\documentclass{pasj02}
\Received{}
\Accepted{}

\usepackage[switch, mathlines]{lineno}

\begin{document}

\title{Quantitative analysis of the molecular gas morphology in nearby disk galaxies}

\author{Takashi Yamamoto,\altaffilmark{1,}$^{*}$
        Daisuke Iono,\altaffilmark{2}
        Toshiki Saito,\altaffilmark{3}
        Nario Kuno,\altaffilmark{1}\\
        Sophia K. Stuber,\altaffilmark{4}
        Daizhong Liu,\altaffilmark{5} and
        Thomas G. Williams \altaffilmark{6}
        }%
\altaffiltext{1}{University of Tsukuba, 1-1-1 Tennoudai, Tsukuba, Ibaraki 305-8577, Japan}
\altaffiltext{2}{National Astronomical Observatory of Japan, 2-21-1 Osawa, Mitaka, Tokyo 181-8588, Japan}
\altaffiltext{3}{Faculty of Global Interdisciplinary Science and Innovation, Shizuoka University, 836 Ohya, Suruga-ku, Shizuoka 422-8529, Japan}
\altaffiltext{4}{Max-Planck-Institut f\"{u}r Astronomie, K\"{o}nigstuhl 17, 69117 Heidelberg, Germany}
\altaffiltext{5}{Purple Mountain Observatory, Chinese Academy of Sciences, 10 Yuanhua Road, Nanjing 210023, China}
\altaffiltext{6}{Sub-department of Astrophysics, Department of Physics, University of Oxford, Keble Road, Oxford OX1 3RH, UK}
\email{yama.boulder@gmail.com}

\KeyWords{galaxies: ISM${}_1$  --- galaxies: structure${}_4$ --- ISM: structure${}_5$}

\maketitle

\begin{abstract}
We present a quantitative and statistical analysis of the molecular gas morphology in 73 nearby galaxies using high spatial resolution CO~(\textit{J}\ =\ 2--1) data obtained from the Atacama Large Millimeter/submillimeter Array (ALMA) by the PHANGS large program. We applied three model-independent parameters:  
Concentration ($C$), Asymmetry ($A$), and Clumpiness ($S$) which are commonly used to parameterize the optical and infrared morphology of galaxies. 
We find a clear apparent correlation between $A$ and $S$, with a Spearman's rank correlation coefficient of $0.52$ with a $p$-value of $2\times10^{-6}$. This suggests a higher abundance of molecular clumps (i.e. giant molecular cloud associations) in galaxies that display stronger distortion or biased large-scale molecular gas distribution. In addition, the analysis of the $C$ parameter suggests high central molecular concentration in most barred spiral galaxies investigated in this study. Furthermore, we found a positive correlation between the length of the bar structure ($R_\mathrm{bar}/R_{25}$) and the $C$ parameter, with a Spearman's rank correlation coefficient of $0.63$ with a $p$-value of $3.8 \times 10^{-5}$, suggesting that larger bar structure can facilitate overall molecular gas transport and yield higher central concentration than galaxies with shorter bars. Finally, we offer a possible classification scheme of nearby disk galaxies which is based on the CAS parameters of molecular gas.

\end{abstract}

\section{Introduction}
The study of galaxy morphology began with Edwin Hubble's qualitative classification ~\citep{Hubble1936}, which remains a widely used method to classify the morphological type of galaxies even in the literature today. Subsequent catalogs by \citet{de Vaucouleurs1959}, \citet{van den Bergh1960}, and \citet{Sandage1961} provide a more detailed morphological classification of nearby galaxies based on quantitative analysis of the stellar light structure obtained from optical images. At the end of the 20th century, a more quantitative approach to analyzing galaxy structure started to gain more attention. Various parameters were used to quantify the galaxy structure, such as the concentration of light \citep{de Vaucouleurs1974}, the index of concentration, rotational symmetry \citep{Schade1995, Abraham1996} or the patchiness of the light distribution ~\citep{Isserstedt1986}.

\citet{Conselice2003} introduced a model-independent quantitative parameters for the morphological classification of galaxies based on three structural indices: Concentration ($C$), the concentration of stellar light, Asymmetry ($A$), the asymmetry of galaxy morphology, and Clumpiness /Smoothness ($S$), representing the degree of luminosity patchiness in a galaxy. Called "CAS" morphological classification, ~\citet{Conselice2003} argues that the CAS system can be used to distinguish stellar light distribution in nearby galaxies based on different phases of evolution as well as the modes of galaxy evolution, such as a major merger. Other methods of measuring concentration include the Gini coefficient ~\citep{Abraham2003} and the second-order moment of the brightest 20\% of the galaxy’s flux $M_{20}$~\citep{Lotz2004}.

The CAS morphological parameters are also used for quantifying and analyzing the structure of high-resolution images taken toward distant galaxies at various wavelengths (e.g.,~\cite{Conselice2007};~\cite{Conselice2014}).
In ultraviolet emission, \citet{Mager2018} found that galaxies tend to appear more late-type when analyzed with the CAS parameters. For example, early type galaxies (elliptical galaxies and lenticular galaxies) generally become less concentrated, more asymmetric, and more clumpy as the observed wavelengths become shorter. In $\mathrm{H \alpha}$ emission, \citet{Fossati2013} found that the CAS parameters of $\mathrm{H \alpha}$ images show no correlation with Hubble type, suggesting that massive galaxies grow inside-out. A recent study by  \citet{Ferreira2022} used the James Webb Space Telescope (JWST) data and classified 280 ($3<z<8$) galaxies into three categories based on their visual morphology: spheroids, disks, and peculiar, finding an overall dominance of disk structure in galaxies observed in rest-frame optical. 

Historically, the CAS morphological classification has been used mainly to analyze the stellar light distribution in galaxies, utilizing the vast availability of a uniform set of high-quality optical images obtained across the sky. Stars are not the only baryons associated with galaxies, and there are many reasons for quantifying the structure of molecular gas, the fuel for star formation since the dynamics and response of the gas are largely affected by the underlying stellar potential and episodic events such as star formation or AGN activities. Several studies in the past have investigated the concentration of CO gas using single dish and interferometric data (e.g.,~\cite{Sakamoto1999};~\cite{Sheth2005};~\cite{Kuno2007}), even though the CAS parameter was not employed. 
In a related context, the CAS parameter was used to analyze the distribution of HI gas, showing that the Asymmetry index of HI gas can be a valuable indicator of galaxy interactions \citep{Holwerda2013}. 
Over the past years, the Atacama Large Millimeter/submillimeter Array (ALMA) has produced a significant set of high-quality images of molecular gas in nearby galaxies, enabling a systematic study of the structure of molecular gas and investigate how it is related to large-scale properties such as bars or the global star-formation activity. 

The CAS analyses have recently been applied to high-quality molecular gas data (e.g.,~\cite{Davis2022};~\cite{Lee2022};~\cite{Roberts2023}). \citet{Davis2022} analyzed the morphology of the molecular gas of 86 nearby galaxies, including early-type galaxies (ETGs), using Asymmetry, Smoothness, and Gini. They suggested that the effective stellar mass surface density is the strongest factor determining the morphology of the molecular gas.  \citet{Lee2022} and \citet{Roberts2023} investigated the environmental effects of galaxies by Asymmetry measurements using CO and HI data compared to the isolated galaxies. The former study is on IC 1459 and NGC 4636 groups, while the latter is on the Virgo cluster galaxies. \citet{Lee2022} showed that the group galaxies tend to have asymmetric CO and HI gas morphology. \citet{Roberts2023} found that the Virgo galaxies have high CO Asymmetry values that are only marginally larger than those of the isolated galaxies, and there is a weak correlation between HI and CO Asymmetry values over their entire galaxies. Furthermore, they showed a stronger correlation for galaxies strongly impacted by environmental perturbations. 

The main goal of this paper is to statistically analyze the molecular gas distribution in nearby galaxies by parameterizing their structure according to the CAS classification scheme. 
We use molecular gas data obtained toward 73 galaxies selected from the archive data of the \underline{P}hysics at \underline{H}igh \underline{A}ngular resolution in \underline{N}earby \underline{G}alaxie\underline{S} (PHANGS) survey with ALMA (PHANGS-ALMA).
PHANGS-ALMA aims to obtain cloud-scale CO~(\textit{J}\ =\ 2--1) maps for all ALMA-visible, massive, star-forming disk galaxies out to the near side of the Virgo Cluster ~\citep{Leroy2021b}. 

This paper is organized as follows: section 2 describes the data for the 73 galaxies used in this paper and explains the CAS parameters. section 3 presents the key results revealed by this analysis, section 4 discusses the results in detail, and section 5 is a summary.

\section{Data and analysis method}

\subsection{PHANGS-ALMA archival data}
We retrieved the PHANGS-ALMA archive data v4.0\footnote{https://sites.google.com/view/phangs/home/data} from the Canadian Advanced Network for Astronomical Research\footnote{https://canfar.net/} (CANFAR) consortium in Canadian Astronomy Data Centre (CADC). 
The survey data were taken as a Large Program in ALMA Cycle 5 (2017.1.00886L, PI: E.Schinnerer). The Public Release contains CO~(\textit{J}\ =\ 2--1) data for 90 massive, star-forming, and close to face-on galaxies visible to ALMA within 20~Mpc, including galaxies characterized by strong bars, grand design, flocculent, and lenticular morphologies. The processing of the PHANGS-ALMA survey and its pipelines are discussed in ~\citet{Leroy2021a}. Of these 90 galaxies, 74 are originally in the main sample \cite{{Leroy2021b}}. The main sample images were constructed with data from the 12-m array, the 7-m array, and the Total Power Array, meaning that all spatial frequencies were observed and recovered.

The PHANGS-ALMA data consists of two types of products: the strict mask products and the broad mask products. We analyzed the molecular gas morphology according to the modified CAS system presented in Section 2.2 using both products, and we find that the results are overall similar. However, the Moment 0 maps of the broad mask products are more complete and cover larger fractions of the total flux (98\%). The completeness of strict mask-to-broad mask is 65\%, and the strict mask products have a loss of 35\% of broad mask flux. Therefore, we have proceeded to use the broad mask products for the analysis in this study.

We selected 73 galaxies by visually inspecting the images and selecting galaxies with sufficient spatial resolution for depicting galaxy structures with no missing parts in the image. NGC~6744 has an incomplete coverage pattern. There are two separated regions to the north and south of the galaxy's bulge. The middle strip has not been observed. Therefore, NGC~6744 image was excluded from this analysis. Our sample includes 19 galaxies from the Virgo Cluster (by the Virgo Cluster Catalogue;~\cite{Binggeli1985}), as well as some galaxies from the Eridanus Cluster, the Fornax Cluster, and the Drado Group. NGC 1300 and NGC 1385 belong to the Eridanus Cluster (\cite{Willmer1989}), NGC 1317 and NGC 1365 belong to the Fornax Cluster (\cite{Morokuma2022}), and NGC 1433 and NGC 1566 belong to the Dorado Group (\cite{Tikhonov2020}). A total of six galaxies are included in clusters and groups. The Dorado is also referred to as a cluster, but it is less concentrated. These clusters are considerably smaller than the Virgo Cluster, with a maximum of around 100 galaxies, making them poor clusters. These galaxies may have experienced gas stripping due to ram pressure or tidal interactions (e.g.,~\cite{Cortese2021}). In such cases, the Asymmetry ($A$) index values may be larger, as shown in~\citet{Lee2022} and~\citet{Roberts2023}.

\begin{figure}[t]
 \begin{center}
  \includegraphics[width=80mm]{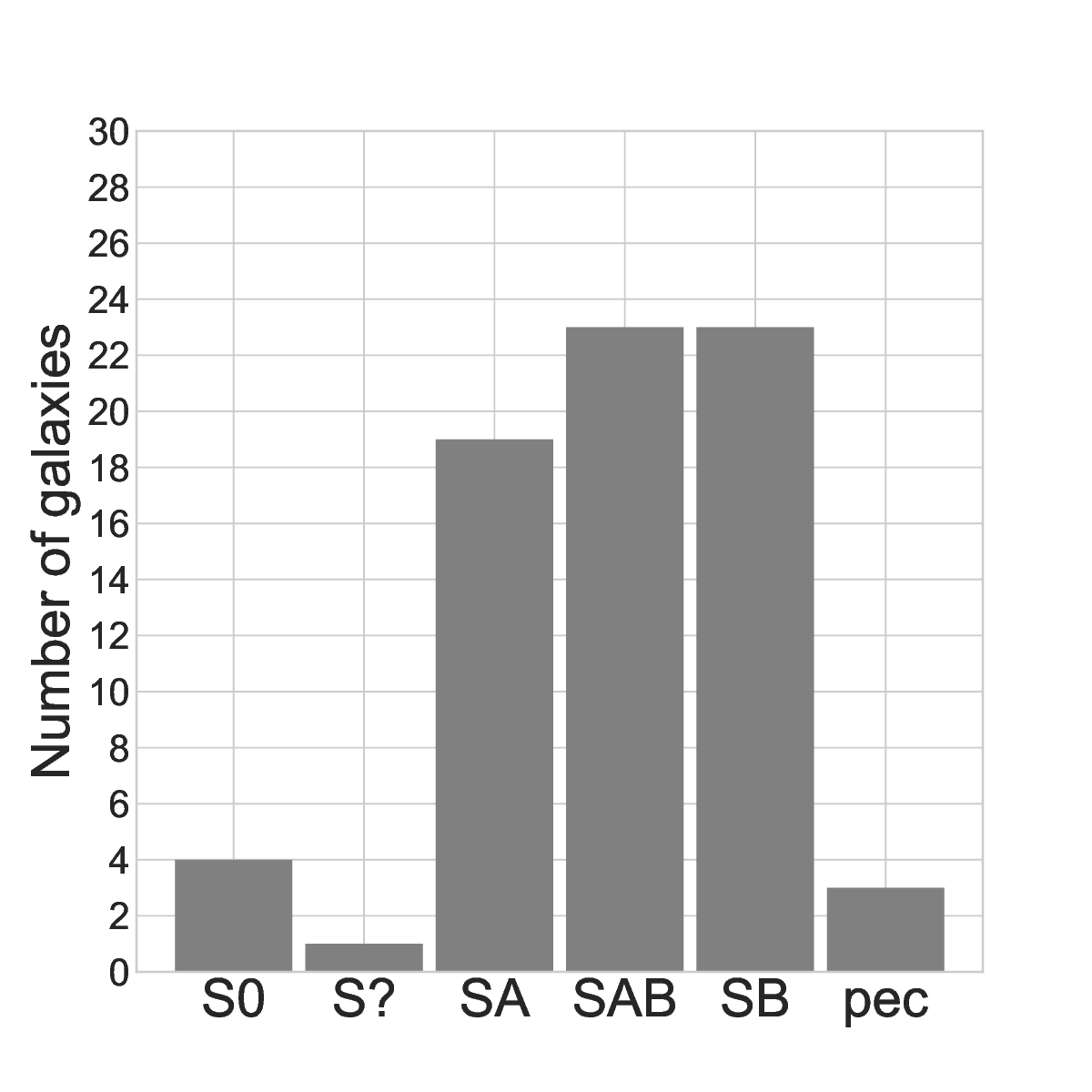}
 \end{center}
 \caption{Number of galaxies by morphological type. Histogram of the morphological type of the sample galaxies. This classification is based on RC3. The S0 galaxy includes both SA0 and SB0. "S?" refers to a galaxy with a doubtful classification as a spiral galaxy.}
 \label{fig:galaxy_type2}
\end{figure}

Figure~\ref{fig:galaxy_type2} shows the distribution of the morphological type according to the  Third Reference Catalogue of Bright Galaxies (RC3;~\cite{RC3}) which is based on optical images.
The majority (65; $90\%$) of the galaxies are normal spirals, within which 48 have bar structures and are classified as SB (strongly barred spirals) or SAB (intermediate barred spirals). The sample only contains four S0 (lenticular galaxies) and three peculiar galaxies, and therefore, the analysis of these galaxies contains uncertainties pertaining to the small sample size.

Our analyses utilize the distribution of the CO (\textit{J}\ =\ 2--1) emission line.  
The apparent similarity in the distribution of CO~(\textit{J}\ =\ 1--0) and CO~(\textit{J}\ =\ 2--1) emission lines seen in past studies (e.g.,~\cite{den Brok2021};~\cite{Yajima2021};~\cite{Egusa2022}) convincingly justifies the use of CO~(\textit{J}\ =\ 2--1) emission line rather than the commonly used CO~(\textit{J}\ =\ 1--0) emission line. 
In addition, although the final PHANGS-ALMA product contains similar angular resolution across the different galaxy sample, the physical resolution can vary as much as a factor of six (30 pc to 180 pc) since the sample contains galaxies at different distances. 
 We convolved all images to the physical resolution of $\sim 180$ pc to ensure a fair comparison across the sample.
Finally, we clipped all data at $0.3\ \mathrm{K\ km\ s^{-1}}$
 which is three times the RMS noise level ($\sigma=0.1\ \mathrm{K\ km\ s^{-1}}$) of the least sensitive data, which was measured in the moment 0 map of NGC~2775.

\subsection{Modified CAS indices system}

The CAS parameters were calculated for each galaxy using a Python script written specifically for this purpose and calling libraries to aid the analysis:  NumPy\footnote{https://numpy.org}~(\cite{Harris2020}), SciPy\footnote{https://scipy.org}~(\cite{Virtanen2020}), Astropy\footnote{https://www.astropy.org}~(\cite{Astropy2018}), and Matplotlib\footnote{https://matplotlib.org}~(\cite{Caswell2021}). The library for advanced image processing OpenCV\footnote{https://opencv.org}~(\cite{Bradski2000}) was used for calculating the Asymmetry index.  

\subsubsection{Concentration ($C$)}

\begin{figure}[t]
  \begin{center}
  \includegraphics[width=80mm]{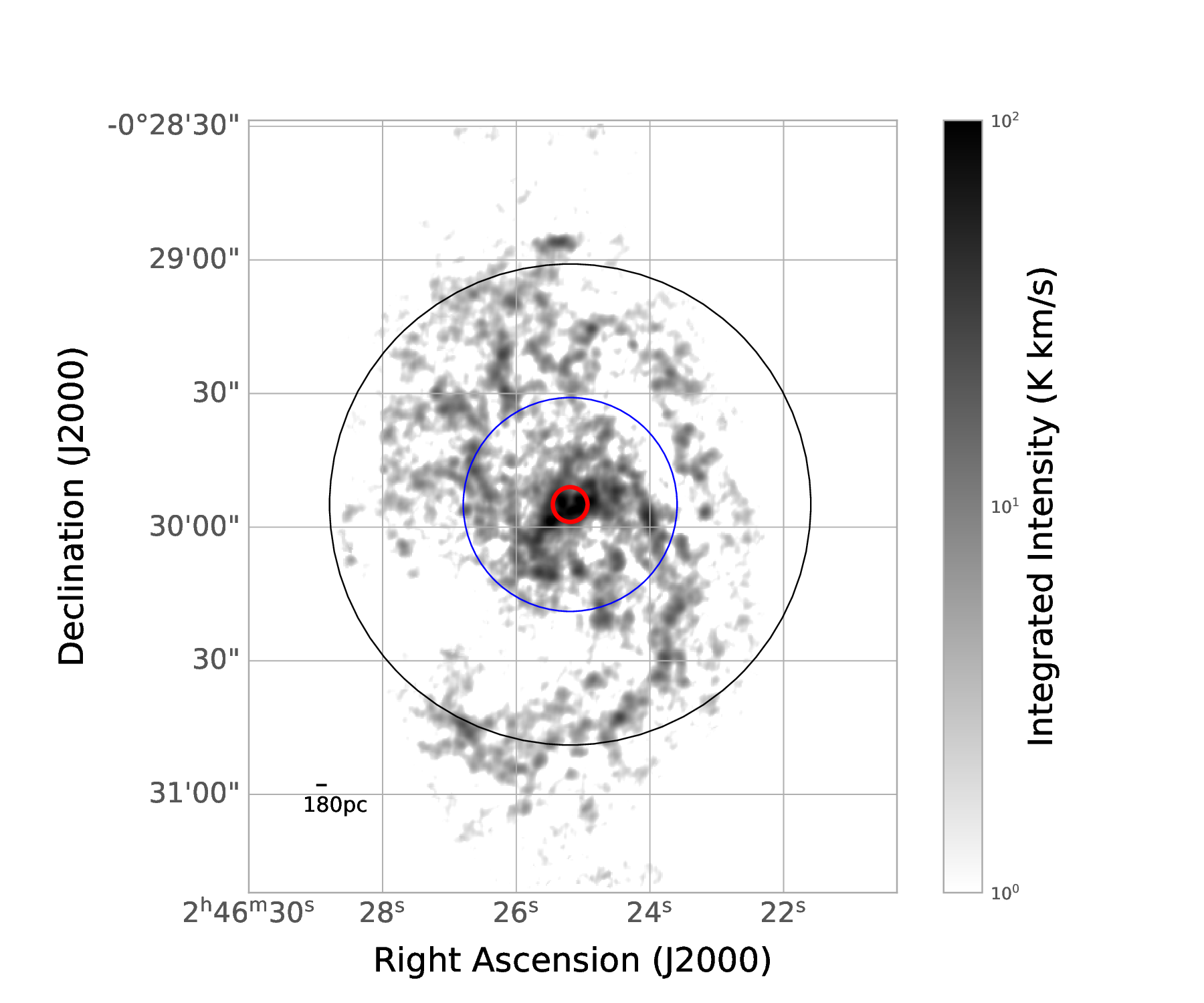}
 \end{center}
 \caption{The Concentration ($C$) index is defined as the ratio of 90\% of total flux to 10\% of total flux ($r_{90}$, $r_{10}$), normalized by logarithm. The circles indicate $r_{90}$ (black circle), $r_{50}$ (blue circle) and $r_{10}$ (red circle) on NGC~1087.}
 \label{fig:C90_10}
\end{figure}

We define the Concentration ($C$) as the logarithm of the ratio of the radius containing 90\% of the total flux ($r_{90\%}$) to the radius containing 10\% of the total flux ($r_{10\%}$). This is expressed as

\begin{equation}
    C = 5 \times \log(r_{90\%}/r_{10\%}),
\label{eq1}
\end{equation}
which is slightly different from the definition adopted by \citet{Conselice2003}, who uses the radius containing 80\% of the total flux ($r_{80\%}$) to 20\% of the total flux ($r_{20\%}$) instead of $r_{90\%}$ to $r_{10\%}$.
We opt to use $r_{90\%}$ and $r_{10\%}$ as the dynamic range of PHANGS-ALMA data is sufficiently high and allows for a better parameterization of the gas concentration (see figure~\ref{fig:C90_10}).

In~\citet{Conselice2003}, all non-parametric parameters, the Concentration ($C$), the Asymmetry ($A$) and the Clumpiness ($S$) were measured within the $1.5\times$ Petrosian radius. However, we did not apply a radius limit in our measurements, as the bulk of molecular gas is concentrated in the galaxy's inner regions. A similar approach is adopted by \citet{Davis2022} in their calculation of $A$ and $S$. 

\subsubsection{Asymmetry ($A$)}

\begin{figure*}[t]
 \begin{center}
  \includegraphics[width=170mm]{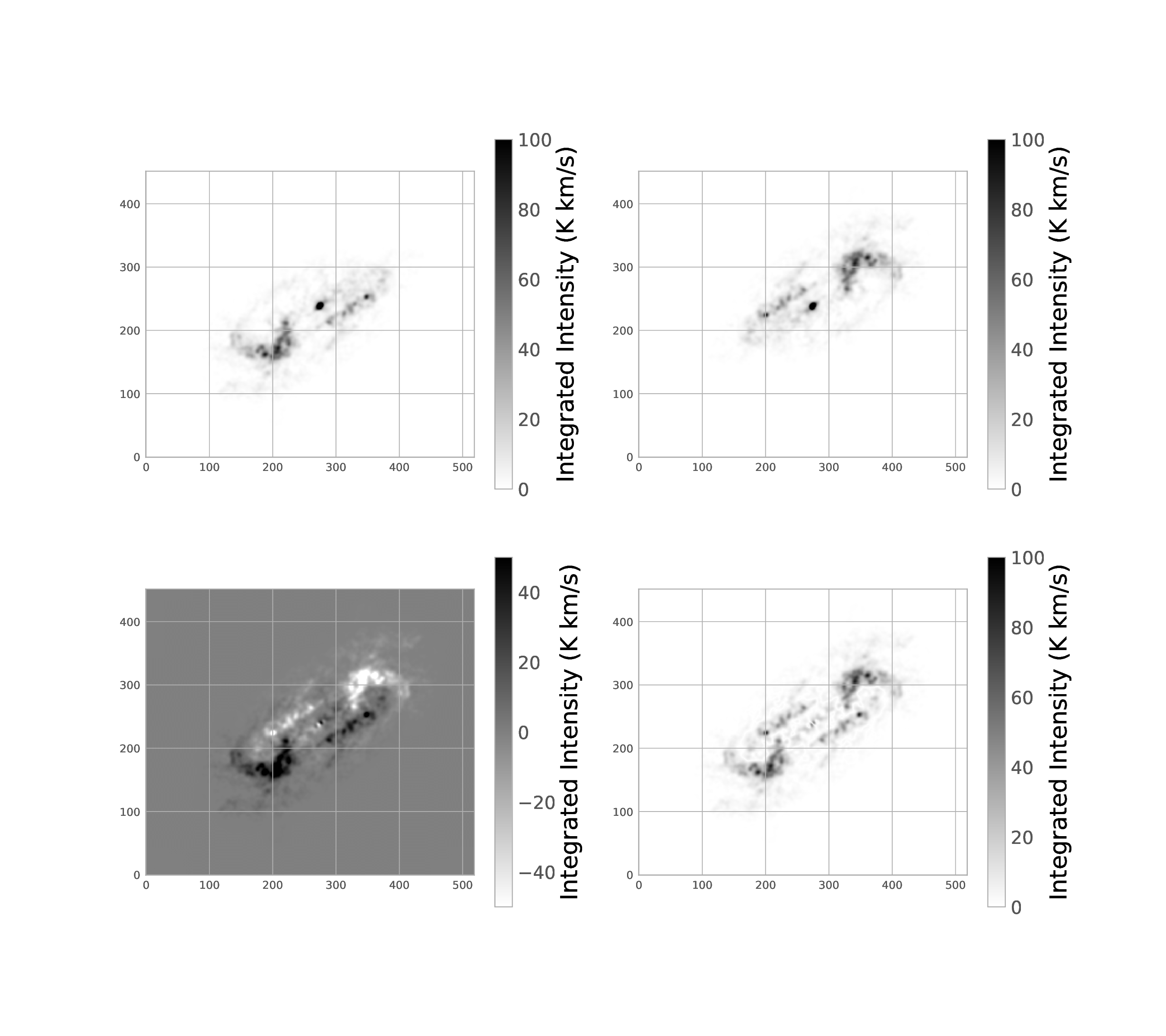}
 \end{center}
 \caption{The calculation process for $A$ in NGC~1511 (classified as SAa(Pec) by RC3). The original image is shown in the top left, while the top right shows the image rotated be 180 degrees around the galactic center $(x,y)=(274,239)$. The bottom left is the residual image, and the bottom right displays the absolute value of the residual. }
 \label{fig:A_index}
\end{figure*}

The Asymmetry ($A$) is computed by rotating the galaxy image 180 degrees around its center and then subtracting it from the original image.  The absolute value of the residuals summed over all pixels is then divided by the total flux of the original image. This is expressed as:

\begin{equation}
        A = \frac{\Sigma_{x,y=1,1}^{M,N}|I_{x,y}-R_{x,y}|}{2\ \Sigma_{x,y=1,1}^{M,N} I_{x,y}},
\label{eq2}
\end{equation}
where $I_{x,y}$ is the original image, and $R_{x,y}$ is the rotated image. $M$ and $N$ are the pixel size of the galaxy image. The minimum value of the asymmetry parameter is 0 (complete symmetry) and the maximum value is 1 (complete asymmetry). By definition, 
$A$ is most sensitive to asymmetric features where the pixel flux is highest. While the central regions of galaxies often exhibit the highest pixel flux, this is not always the case. In some instances, the highest pixel flux may be found in the outer regions of galaxies.

The determination of the galaxy center is crucial for a reliable measurement of the $A$. In this study, we utilized table 3 of ~\citet{Leroy2021b}. They first adopt photometric centers from ~\citet{Salo2015} if available, as these centers were determined by sensitive $3.6\ \mathrm{\mu m}$ near-infrared imaging and should accurately reflect the center of the galaxy's stellar mass. If photometric centers are not available, they adopt central positions from the 2MASS Large Galaxy Atlas ~\citep{Jarrett2003}. As a last resort, they use the optically defined central position from HyperLEDA or NED databases. 
In our visual inspection of the PHANGS-ALMA image 
of NGC 1511, we observed a bright 
molecular concentration at the center, 
which likely represents the galaxy’s nucleus. 
Given that this concentration is offset from the coordinates used by \citet{Leroy2021b}, we have designated the peak of this molecular concentration as the center of NGC~1511 ($\alpha\ (J2000) = 3:59:36.45,\ \delta\ (J2000) = -67:38:01.56$). Finally, the choice of photometric center generally only affects the calculation by a few arcseconds or less, and therefore, a fiducial uncertainty of $1''$ was adopted.

Figure~\ref{fig:A_index} illustrates the process of calculating $A$. The interacting galaxy NGC~1511  (e.g.,~\cite{Koribalski2005}) is classified as SAa (Pec) by RC3 and has a high $A$ value of 0.70. However, even non-interacting galaxies can exhibit high $A$ values when analyzed in the context of their CO~(\textit{J}\ =\ 2--1) gas distributions. Finally, the error associated with the $A$ was estimated by shifting the center coordinates by the beam size and quantifying the range of $A$ obtained through this process. 

\subsubsection{Clumpiness ($S$)}

\begin{figure*}[ht]
 \begin{center}
  \includegraphics[width=140mm]{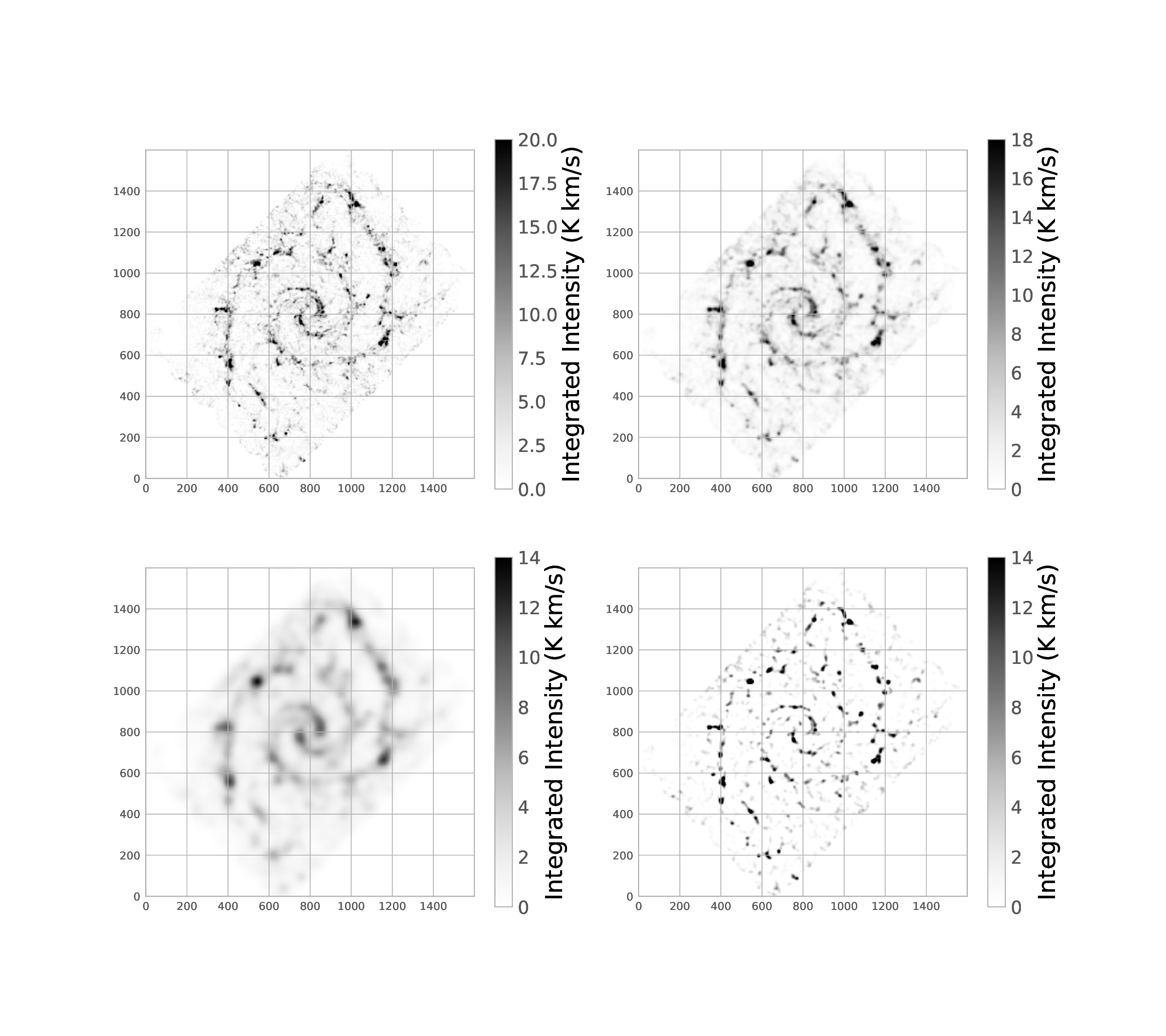}
 \end{center}
 \caption{The calculation process of the Clumpiness ($S$) index for the galaxy NGC~628. The top left panel shows the original PHANGS-ALMA image with a physical resolution of $\sim54\ \mathrm{pc}$. The top right panel displays the image with a resolution of $180\ \mathrm{pc}$ ($I_{x,y}$), while the bottom left panel displays the image with the smoothing filter $\sigma$ applied ($I_{x,y}^\sigma$). Here, we adopt $\sigma=500 \mathrm{pc}$. The bottom right panel presents the residual image, highlighting the presence of large clumps of GMAs.
 }
 \label{fig:S_index}
\end{figure*}

\begin{figure}
 \begin{center}
  \includegraphics[width=95mm,height=80mm]{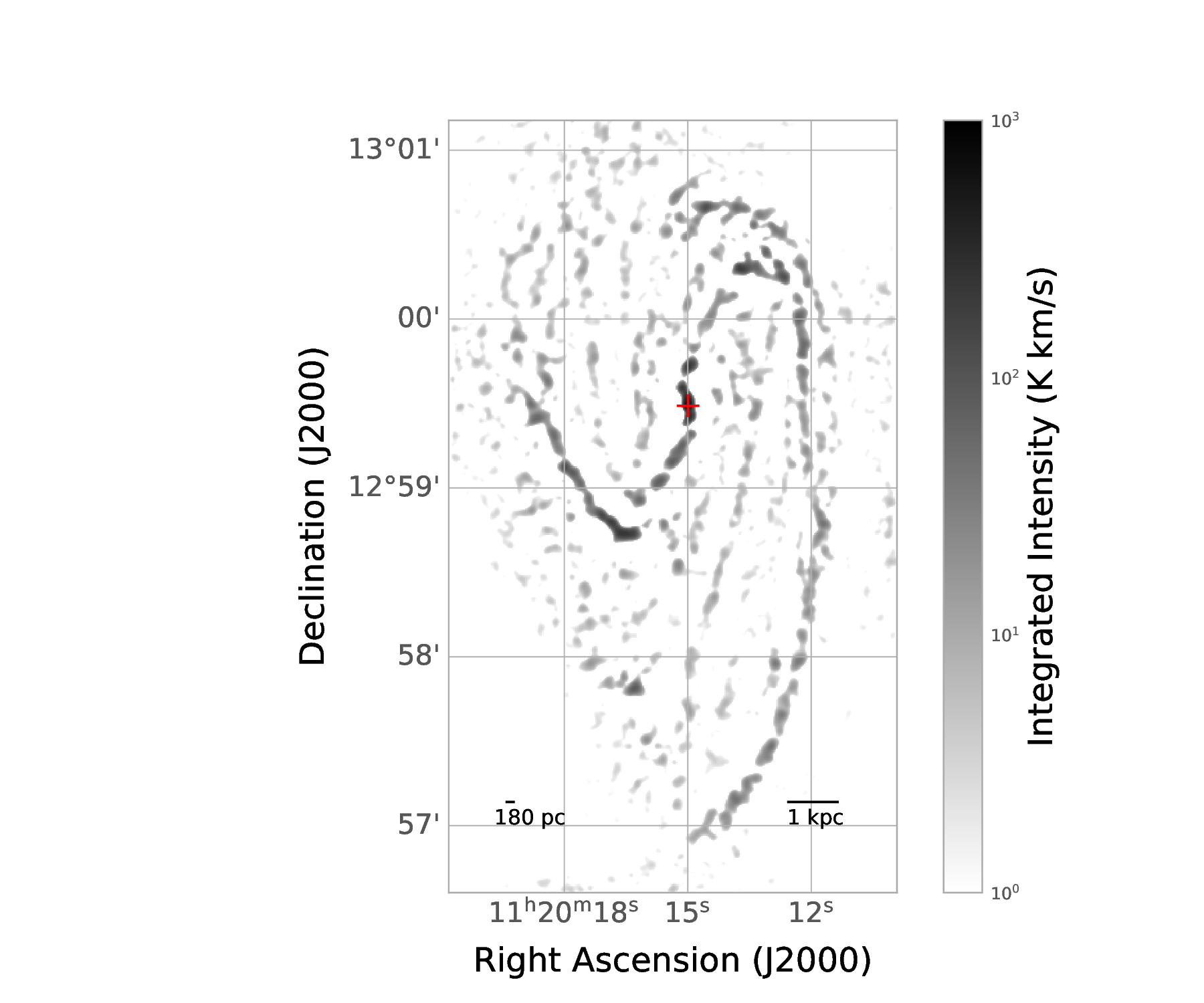}
  \end{center}
 \caption{The CO~(\textit{J}\ =\ 2--1) intensity image used for Clumpiness calculation of NGC~3627. The red cross indicates the central position of this galaxy. }
 
 \label{fig:S_ngc3627}
\end{figure}

Clumpiness ($S$) is an index of the patchiness of light distribution in a galaxy \citep{Conselice2003}, and the original definition is;

\begin{equation}
    S=10\times \Sigma_{x,y=1,1}^{M,N} \frac{(I_{x,y}-I_{x,y}^\sigma)-B_{x,y}}{I_{x,y}},
\label{eq3}
\end{equation}
where $I_{x,y}$ is the flux value of the galaxy at position $(x,y)$, $I_{x,y}^\sigma$ is the same value obtained after reducing in resolution by a smoothing filter with width $\sigma$, $B_{x,y}$ is the background value at $(x,y)$, and $M$ and $N$ are the pixel size of the galaxy image. 
The spatial resolution of $I_{x,y}$ is 180~pc and enables us to identify giant molecular cloud associations (GMAs) larger than that scale. 
The $S$ index measures the ratio of the distribution of GMAs to the overall molecular gas distribution in the galaxy.

The definition of $S$ adopted in this study, where the background value is zero, is given by the following equation:

\begin{equation}
    S=10\times \Sigma_{x,y=1,1}^{M,N} \frac{I_{x,y}-I_{x,y}^\sigma}{I_{x,y}},
\label{eq4}
\end{equation}
where the same definition from equation~(\ref{eq3}) for $I_{x,y}$ and $I_{x,y}^\sigma$ are adopted.

The final value of $S$ can be sensitive to the adopted value of $\sigma$. Here, \citet{Conselice2003} used $\sigma = 0.3\times r$ (the Petrosian radius), but we used $\sigma = 500 \mathrm{pc}$ as in \citet{Davis2022}.
Figure~\ref{fig:S_index} shows the process of calculating the $S$ in NGC~628, demonstrating that $S$ is a reasonable indicator tracing the clumpiness and smoothness of GMAs at a size scale of $\sim 200\mathrm{pc}$. We note that the close similarity between the $S$ image presented here (see bottom right of figure~\ref{fig:S_index}; figure~\ref{fig:S_ngc3627}) and figure~2 of \citet{Rosolowsky2021} suggests that difference in resolution and adopted method are relatively insignificant.

We argue that the use of the Gini coefficient must be treated with caution when used for quantifying the central concentration of gas. The Gini coefficient loses information on the position coordinates. For example, in the case of molecular gas distribution with a central hole and strong ring structure, the Gini coefficient would result in a high concentration value despite the apparent low concentration of gas in the center (see Figure 7. in~\cite{Davis2022}). Instead, the Gini coefficient measures the equality of gas distribution and serves as an alternative to Clumpiness ($S$), which measures the smoothness of the gas distribution \citep{Davis2022}. 

Finally, no correlations were observed between the CAS morphological parameters and the inclination angle of the galaxies, and therefore, we did not correct the inclination angle throughout this paper. The inclination angle ($i$) of the sample are all less than $75^\circ$.

\section{Results}
The results of the CAS calculations for all galaxies are presented in table~\ref{tab:001}. 
In table~\ref{tab:002}, we show the mean, median, and standard deviations associated with each parameter. The physical properties of the PHANGS-ALMA galaxy are in table~\ref{tab:003}. We investigate the relation among the three structural parameters in what follows.

\subsection{Correlation between Asymmetry and Clumpiness}

The strongest correlation is seen in the relationship between $A$ and $S$ (see figure~\ref{fig:A1_S}; table~\ref{tab:004}), with a Spearman's rank correlation coefficient of 0.52 and the associated $p$-value of $2\times 10^{-6}$.
A recent study \citep{Davis2022} used the millimetre-\underline{W}ave \underline{I}nterferometric \underline{S}urvey of \underline{D}ark \underline{O}bject \underline{M}asses (WISDOM) and PHANGS data taken by ALMA to analyze correlations between galaxy center's Asymmetry, Smoothness (Clumpiness), and the Gini coefficient with galaxy stellar mass and star formation efficiency (SFE), among other factors. \citet{Davis2022} showed that smooth interstellar mediums (ISMs) are symmetric while clumpy ISMs are asymmetric; They showed that Spearman's rank correlation coefficient between clumpiness and asymmetry was 0.72, with a $p$ value of $2\times 10^{-15}$. Our results show a qualitatively similar trend to~\citet{Davis2022}. The stronger correlation of~\citet{Davis2022} might be due to differences in the data sample, as the inclusion of WISDOM early-type galaxies (ETGs) in ~\cite{Davis2022} may have contributed significantly (See Figure 2. in~\cite{Davis2022}).

It is possible that the asymmetric distribution of local clumps across the galaxy disk increases $A$, amplifying the correlation between $A$ and $S$. To test how the small-scale asymmetry due to locally condensed clumps affects the value of $A$, we subtracted the value of each pixel of $S$ from the value of the same pixel in the original image and re-calculated $A$ (called $A^{\prime}$). The correlation between $A^{\prime}$ and $S$ is shown in figure~\ref{fig:A_S}. The Spearman's rank correlation coefficient is 0.44 with an associated $p$-value of $1\times 10^{-4}$, which are similar to those derived for the correlation between $A$ and $S$. This suggests that the correlation between $A$ and $S$ is likely reflecting the global distribution rather than the asymmetries caused by the distribution of the local clumps.

One would naively hypothesize that repeated close interactions and gas stripping, such as those seen in a cluster environment, can increase the asymmetry in the molecular gas distribution, possibly affecting the relationship between $A$ and $S$. Despite this prediction, we find no statistically significant differences in the relationship between $A$ and $S$ when Virgo Cluster, poor clusters (Fornax Cluster, Eridanus Cluster and Dorado group) and non-cluster galaxies are compared (see figure~\ref{fig:A_S_VC}). There is also no significant difference in $A$ between all cluster galaxies and the other galaxies (see figure~\ref{fig:A_VC}). We further ran the Kolmogorov-Smirnov (KS) test on $A$ to check the probability that the two populations, all cluster galaxies (Virgo Cluster and poor clusters) and non-cluster galaxies, are drawn from the same parent sample. We obtain the KS probability value ($p$) of $0.83$ ($> 0.05$), suggesting that $A$ values of the two samples are drawn from the same parent sample. The median $A$ values for all cluster galaxies and non-cluster galaxies are 0.50 and 0.46, respectively. Similarly, the median 
$S$ values are  2.32 for all cluster galaxies and 2.26 for non-cluster galaxies. These median values are exactly the same as for Virgo Cluster and non-Virgo Cluster. The KS test on $S$ returned with $p=0.90$ ($> 0.05$). This result suggests that the two sample $S$ values are also drawn from the same parent sample, further suggesting that clump formation is less affected by frequent dynamical interactions, at least in these particular clusters environment. This is thought to be due to the characteristics of the ISM's distribution, in which molecular gas is abundant in the inner regions of galaxies. In contrast, HI gas is abundant in the outer areas.

This work is largely consistent with results of \citet{Roberts2023}, which is one of the Virgo Environment Traced In CO (VERTICO) collaboration papers, utilizing ALMA Atacama Compact Array (ACA) for CO and Very Large Array (VLA) for HII  observation data. They found that, unlike the HI gas Asymmetry, the CO gas does not take on a significantly larger Asymmetry value even in galaxies where gas stripping occurs.

\subsection{Central concentration of molecular gas in barred spiral galaxies}

We find that the median $C$ values for SB and SAB galaxies are $5.6$ (with the median absolute deviation of $1.6$) and $5.6$ (with the median absolute deviation of $0.9$), respectively. This suggests that the concentration of the two populations is fundamentally the same. Subsequently, we categorized galaxies into two groups: barred spiral galaxies (SB + SAB) and non-barred spiral galaxies (SA). The median $C$ value for barred spiral galaxies (SB + SAB) is $5.6$ (with a median absolute deviation of $1.4$) whereas the median $C$ value for non-barred spiral galaxies (SA) is $3.9$ (with a median absolute deviation of $0.7$). We obtain the KS probability value of $8.6 \times 10^{-6}$, suggesting that $C$ values of the two samples, namely the SA and SB + SAB, are drawn from a different probability distribution. These results further suggest a trend that barred spiral galaxies exhibit higher gas concentration overall. This is also illustrated in the histogram (figure~\ref{fig:hist-C}) and the violin plots (figure~\ref{fig:Violin_C}). 
This result is consistent with \citet{Sakamoto1999}, who reported that molecular gas is more concentrated in the central kilo-parsec in barred galaxies than in non-barred galaxies from CO~(\textit{J}\ =\ 1--0) imaging survey of the central regions of 20 nearby spiral galaxies. Finally, to confirm that our definition of $C$ (see Section 2.1.1) is not producing any biases with respect to that defined by \citet{Conselice2003}, we derived $C$ using the \citet{Conselice2003} definition and plotted in figure~\ref{fig:hist-C82} and figure~\ref{fig:Violin_C82}. As seen in these figures, the difference between the two definitions is negligible.

The scatter plots for the relationship between $C$-$A$ and $C$-$S$ are shown in figure~\ref{fig:A1_C} and  figure~\ref{fig:S_C}, respectively. In these figures, non-barred spiral galaxies (SA), barred spiral galaxies (SB + SAB) and lenticular galaxies (S0) occupy distinct positions in the CAS plane (see section 4.3). We further compare the Virgo cluster galaxies and non-Virgo galaxies in the same parameter space (figure~\ref{fig:A1_C_VC} and figure~\ref{fig:S_C_VC}). Similar to $A$ and $S$ (Section 3.1), there is no significant difference in $C$ between the Virgo cluster galaxies and non-Virgo galaxies, with a median $C$ value of $5.1$ and $4.7$, respectively. The result of KS test is $p=0.92$ ($> 0.05$).

From a detailed examination of figure~\ref{fig:hist-C}, it becomes evident that not all barred spiral galaxies exhibit high concentrations, 
with a subset of barred galaxies with elevated $C$ values ($C \geq 4.9$) and those with $C$ values comparable to non-barred spirals ($3.0 \leq C < 4.9$). The median value of $C$ is 4.9 in all samples.
We note three particular aspects that are observed among barred galaxies which display low $C$ values ($C < 4.9$). Firstly, those with low $C$ values frequently display high $A$ values ($A > 6.0$; see figure~\ref{fig:A1_C} ). 
Secondly, three barred spiral galaxies (NGC~4496A, NGC~4731, NGC~5068)  with especially small stellar masses ($\lesssim 10^{9.5}\  M_{\odot}$) in this sample, may have difficulty concentrating molecular gas in the center. On the relatively high mass side ($>10^{10.5}\  M_{\odot}$), no barred spiral galaxies have $C$ values lower than 4.9. Although only three samples are used here, as discussed below, an intermediate correlation between $C$ and the stellar mass is found in barred spiral galaxies.
The third aspect is the absence of a discernible bar structure in certain galaxies, such as NGC~3521 and NGC~3596. Despite being classified as SAB galaxies in the RC3 catalog, they visually appear to be consistent with SA galaxies and are also categorized as non-barred galaxies in a recent analysis (\cite{Buta2015}; \cite{Querejeta2021}; \cite{Stuber2023}). 
The differences in molecular gas concentrations in barred spiral galaxies will be discussed in more detail in section 4.2.

We note a moderate positive correlation between stellar mass and $C$ (figure~\ref{fig:M_st vs. C}) with a Spearman's rank correlation coefficient of 0.42 and a $p$-value of $2\times10^{-4}$. Limiting the sample to barred spiral galaxies alone gives a Spearman's rank correlation coefficient of 0.39, with a $p$-value of $8\times10^{-3}$, while a similar analysis with just the non-barred spiral galaxies yields a Spearman's rank correlation coefficient of -0.19, and a $p$-value of 0.42. The observed relatively strong correlation in barred spiral galaxies may suggest that the gravitational potential created by stars augments the gravitational influence on the inflowing gas, in addition to the torque applied by the stellar bar structure. This is in contrast to the negligible correlation seen in non-barred spirals.

\subsection{Relationship between CAS parameters and star formation efficiency}

We investigate the relationship between the morphological parameters and other physical quantities such as stellar mass, specific star formation rate (sSFR), and star formation efficiency (SFE)\ (see table~\ref{tab:003}).  The SFE is defined as
\begin{equation}
    \mathrm{SFE(yr^{-1})} = \frac{\mathrm{SFR}}{M_{\mathrm{H_2}}},
\label{eq5}
\end{equation}
where SFR is the star formation rate in $M_{\odot}$~yr$^{-1}$.
The $\mathrm{H_2}$ gas mass ($M_{\mathrm{H_2}}$) is calculated as 
\begin{equation}
    M_{\mathrm{H_2}} = \alpha_{\mathrm{co}} L_{\mathrm{co}},
\label{eq6}
\end{equation}
 where $\alpha_{\mathrm{co}}=4.35\ \mathrm{M_{\odot}} (\mathrm{K\ km\ s^{-1}\ pc^2})^{-1}$ (\cite{Bolatto2013};~\cite{Leroy2021b}) and $L_{\mathrm{co}}$ ($\mathrm{K\ kms^{-1}\ pc^2}$) is the luminosity of the CO~(1--0) line. 
Following the approach taken by ~\citet{Leroy2021b}, the CO~(\textit{J}\ =\ 2--1) to CO~(\textit{J}\ =\ 1--0) ratio of $R_{21}=0.65$ was used to calculate $L_{\mathrm{co}}$. 
The specific star formation rate (sSFR) is defined as

\begin{equation}
    \mathrm{sSFR(yr^{-1})} = \frac{\mathrm{SFR}}{M_{\star}},
\label{eq7}
\end{equation}
where $M_{\star}$ is the stellar mass in $M_{\odot}$.

We analyzed the correlation between $\mathrm{log_{10}SFE}$ and the CAS parameters (see figure~\ref{fig:SFE_S} and figure~\ref{fig:SFE_A}). The physical values associated with each galaxy are summarized in table~\ref{tab:003}. Spearman's rank correlation coefficient between $\mathrm{log_{10}SFE}$ and $A$ is 0.45 with the associated $p$-value of $7\times 10^{-5}$. This result is consistent with the findings of \citet{Davis2022} (Spearman's rank correlation coefficient is 0.44, with the associated $p$-value of $2\times 10^{-5}$). One might argue that the $S$ is an indicator of GMC formation and hence subsequent star formation, in which case we expect a significant correlation between $\mathrm{log_{10}SFE}$ and $S$. Despite this prediction, the correlation between these two quantities appears weakly significant, with a Spearman's rank correlation coefficient of $0.33$ with a $p$-value $5\times 10^{-3}$. The correlation coefficient value for the same correlation derived by \citet{Davis2022} is $0.39$ with a $p$-value $2\times 10^{-4}$.
\section{Discussion}

\subsection{Correlation between Asymmetry and Clumpiness}
\begin{figure}
 \begin{center}
  \includegraphics[width=82mm,height=65mm]{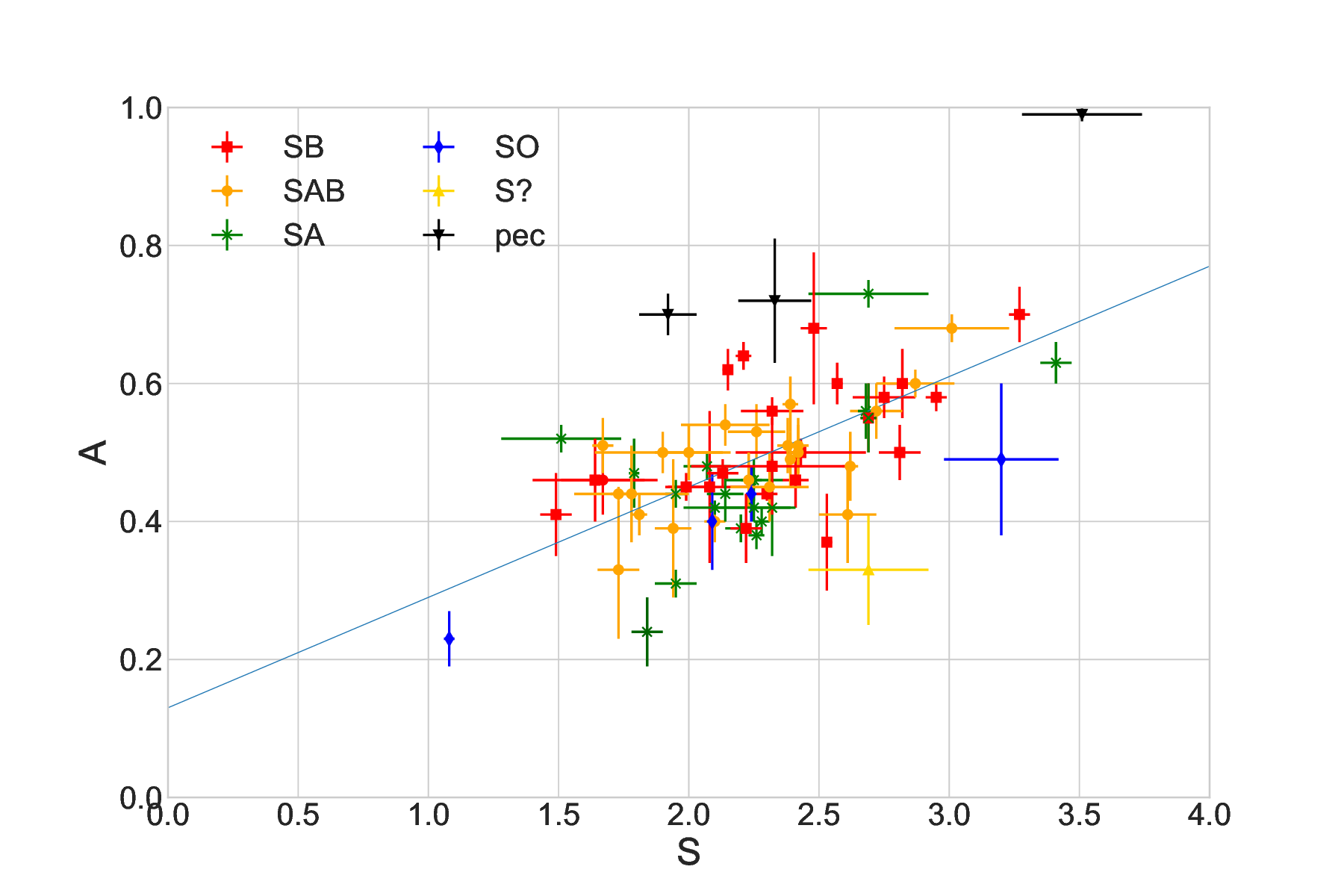}
 \end{center}
 \caption{Scatter plot between the derived $S$ and $A$. SA galaxies are shown as green crosses, SAB as orange circles, SB as red squares, S0 (lenticular) galaxies as blue diamonds, a S? galaxy as gold triangle and peculiar types as black inverted triangles. The sloping thin dotted line is the regression line by the least-squares method.}
 \label{fig:A1_S}
\end{figure}

\begin{figure}
 \begin{center}
  \includegraphics[width=82mm,height=65mm]{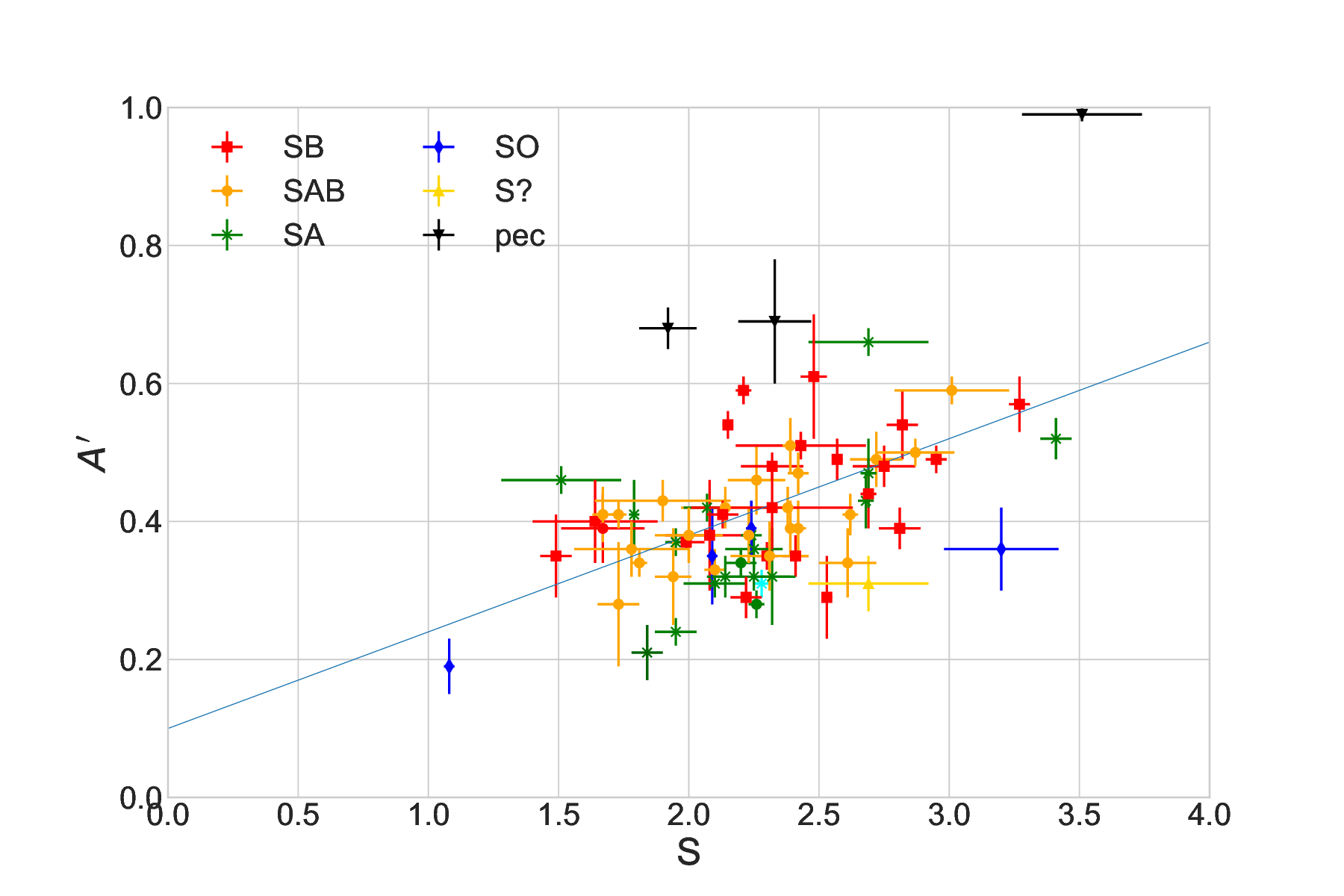}
 \end{center}
 \caption{
 Scatter plot between $S$ and $A^{\prime}$, where $A^{\prime}$ is Asymmetry calculated from the original data minus $S$.}
 \label{fig:A_S}
\end{figure}

\begin{figure}
 \begin{center}
  \includegraphics[width=82mm,height=65mm]{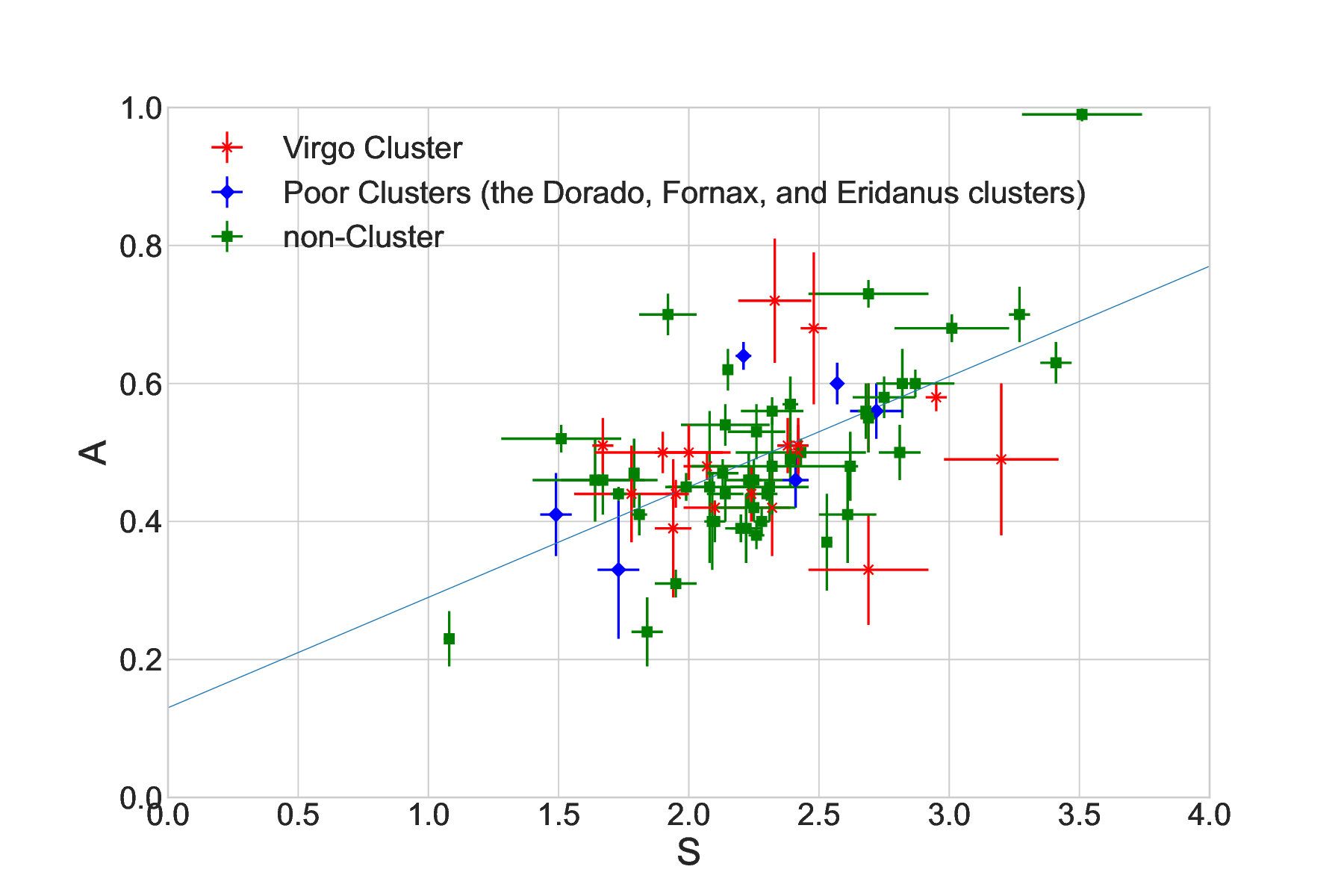}
 \end{center}
 \caption{
 Scatter plot between $S$ and $A$. The Virgo Cluster galaxies are shown as red crosses, poor cluster galaxies as blue diamonds, non cluster galaxies as green squares.}
 \label{fig:A_S_VC}
\end{figure}

\begin{figure}
 \begin{center}
  \includegraphics[width=82mm,height=82mm]{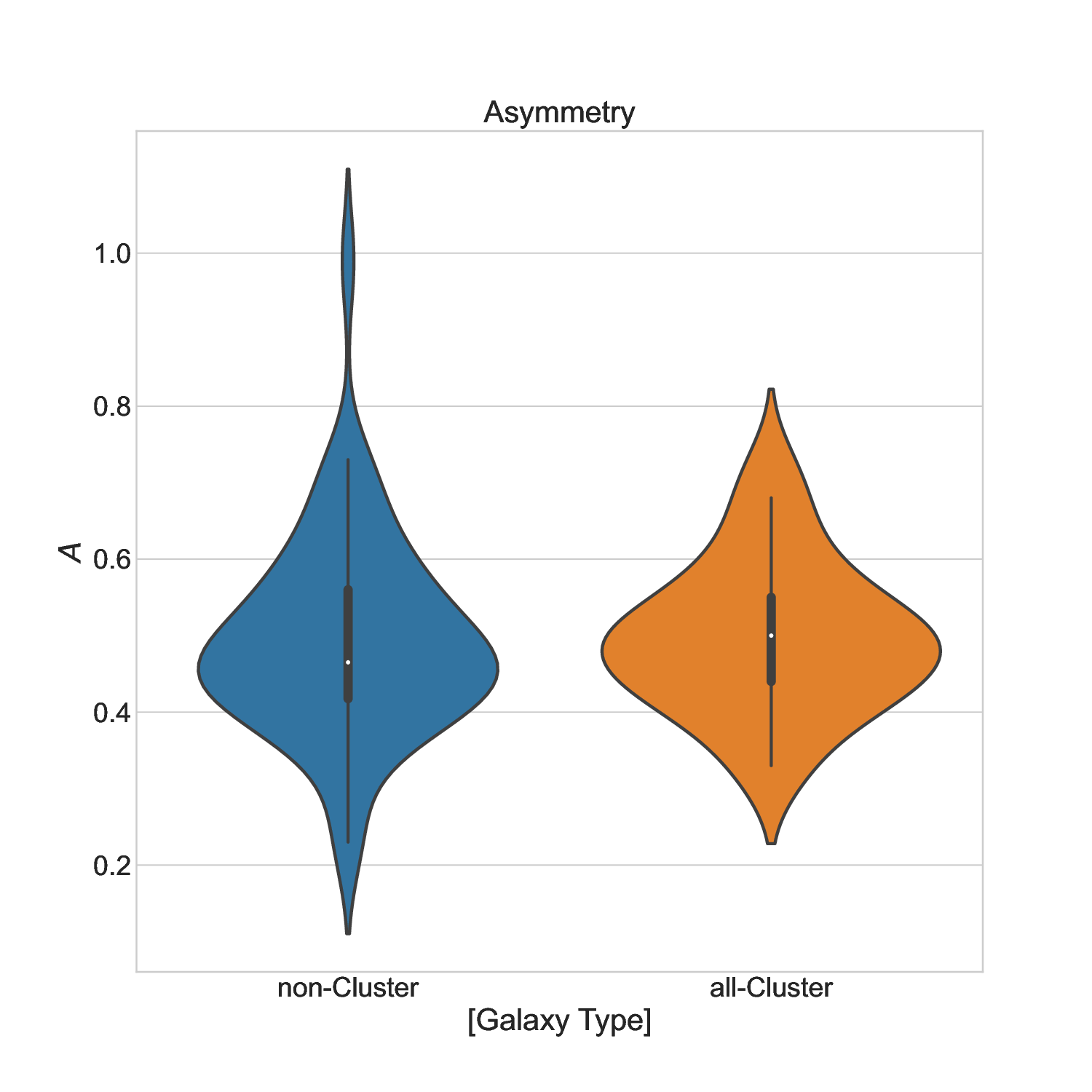}
 \end{center}
 \caption{
 Violin plot of Asymmetry ($A$) between all cluster galaxies and non cluster galaxies.}
 \label{fig:A_VC}
\end{figure}

\begin{figure}
 \begin{center}
  \includegraphics[width=75mm]{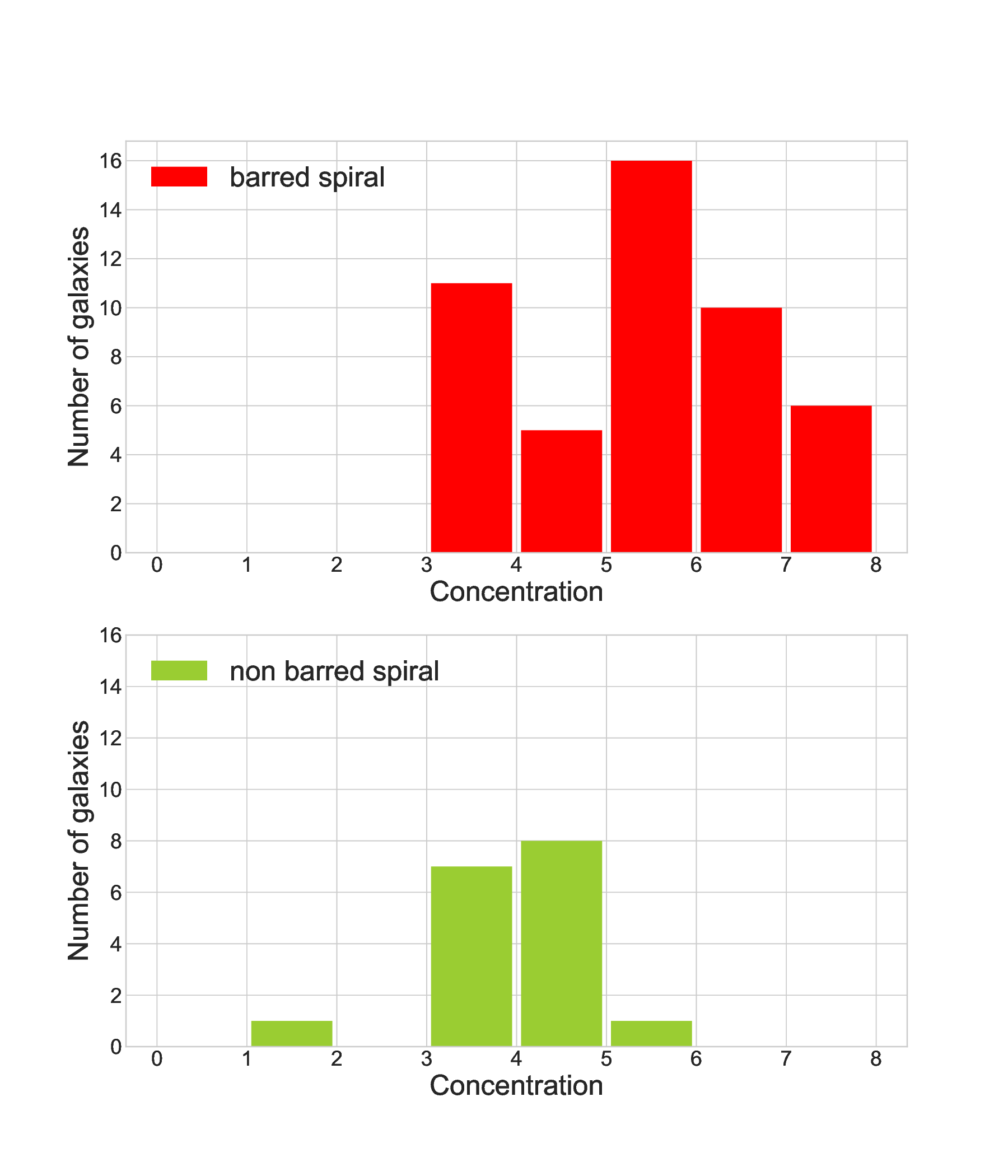}
 \end{center}
 \caption{The histogram of Concentration ($C$) in barred spiral galaxies (the top panel) and non-barred spiral galaxies (the bottom panel). It is clear that galaxies with bar structures tend to be more centrally concentrated.}
 \label{fig:hist-C}
\end{figure}

\begin{figure}
 \begin{center}
  \includegraphics[width=80mm]{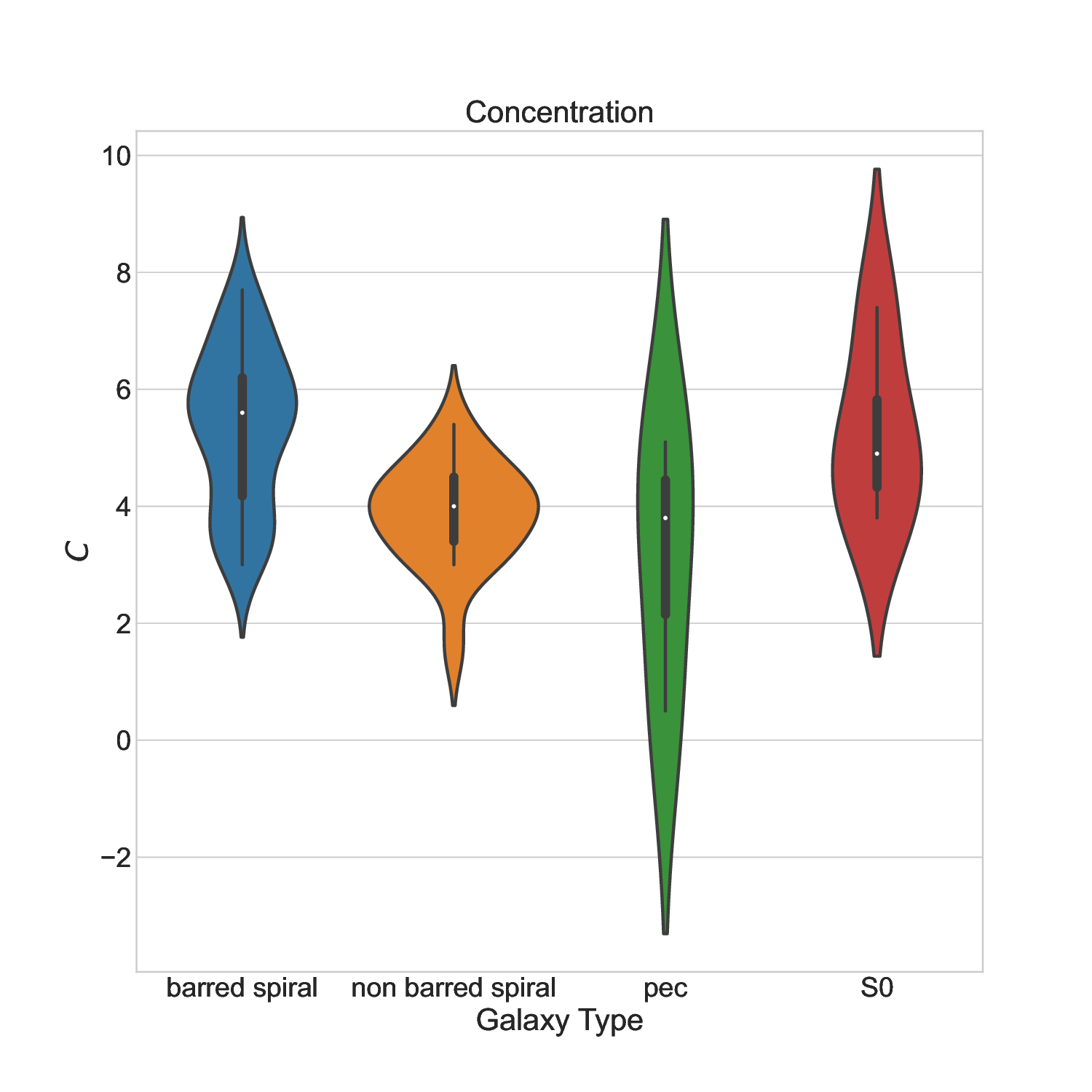}
 \end{center}
 \caption{Violin plot of Concentration by galaxy types. From the left, the order is barred spiral galaxy, non-barred spiral galaxy, peculiar galaxy, and S0 galaxy.}
 \label{fig:Violin_C}
\end{figure}

\begin{figure}
 \begin{center}
  \includegraphics[width=82mm,height=82mm]{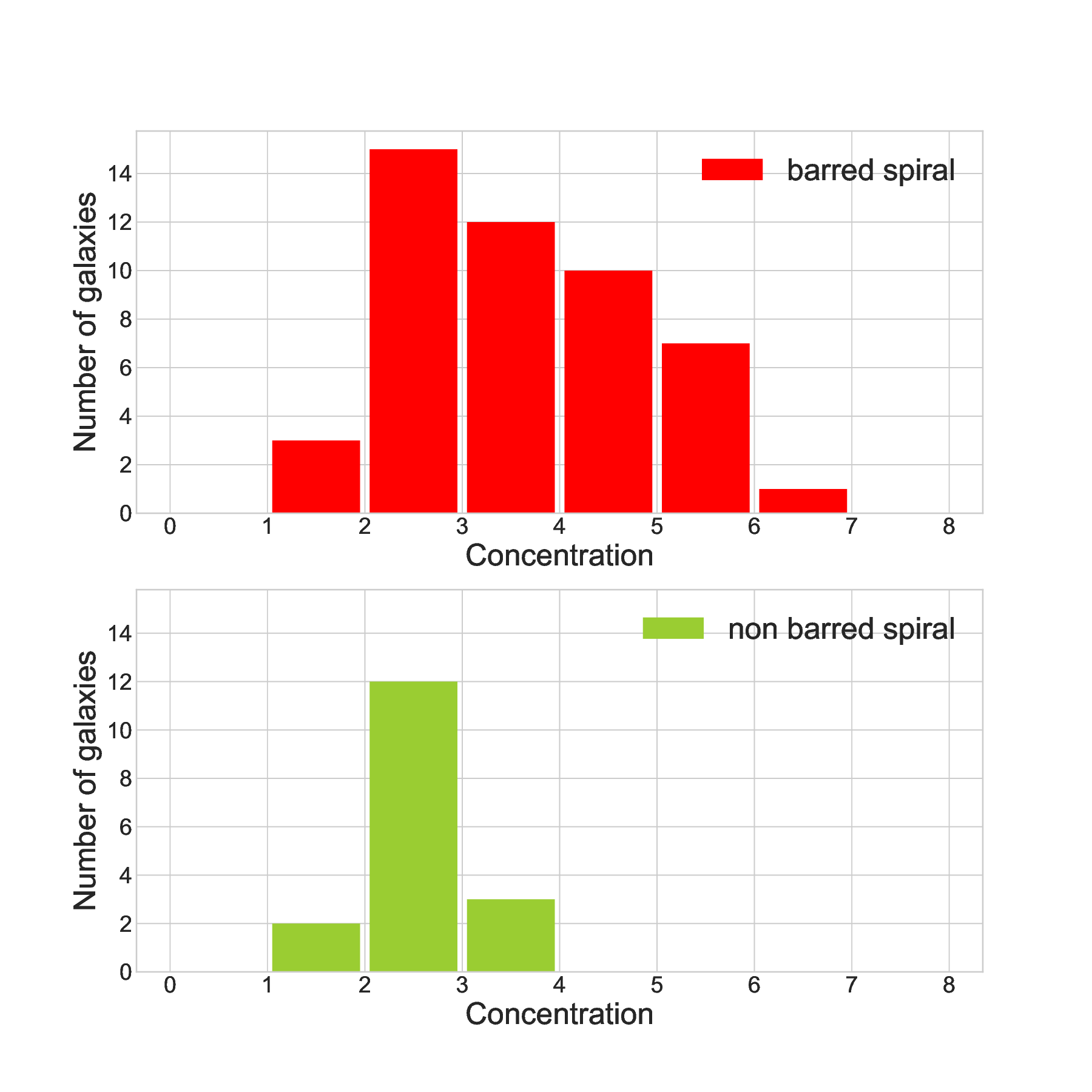}
 \end{center}
 \caption{The histogram of Concentration [$C(80/20)$] in barred spiral galaxies (the top panel) and non-barred spiral galaxies (the bottom panel). It is clear that galaxies with barred structures tend to be more centrally concentrated, as is the case with $C(90/10)$.}
 \label{fig:hist-C82}
\end{figure}

\begin{figure}
 \begin{center}
  \includegraphics[width=80mm]{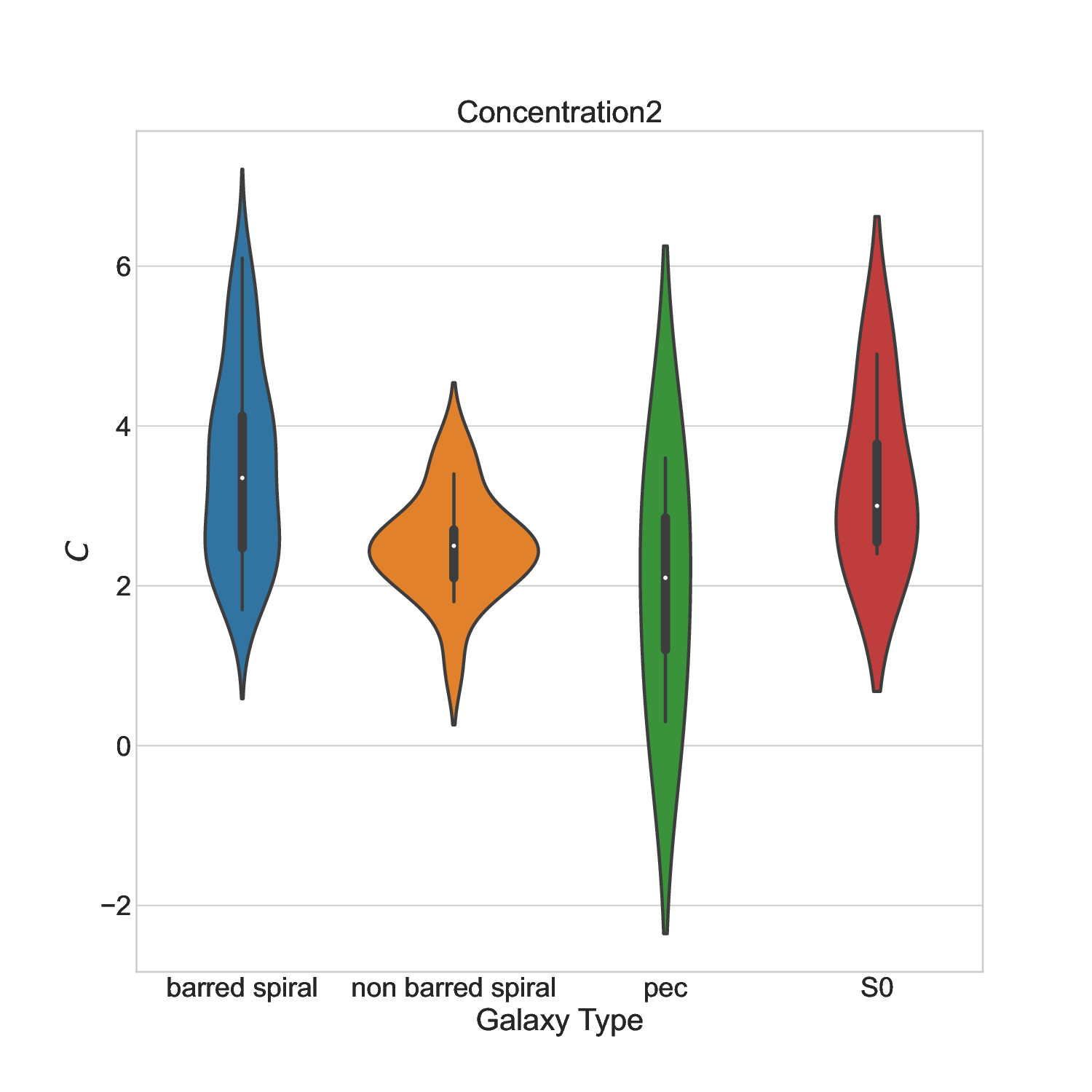}
 \end{center}
 \caption{Violin plot of Concentration [$C(80/20)$] by galaxy types. From the left, the order is barred spiral galaxy, non-barred spiral galaxy, peculiar galaxy, and S0 galaxy.}
 \label{fig:Violin_C82}
\end{figure}

The correlation between $A$ and $S$ presented in section 3.1 naively suggests that morphologically disturbed galaxies are preferentially rich in clumpy structure, since $S$ is a measure of the relative abundance of GMAs, while $A$ is an indicator of a biased or distorted CO gas distribution. 
As the molecular gas mass fraction of the PHANGS galaxies is relatively insignificant of order 10\%, the large-scale gas distribution is primarily dictated by the gravitational potential created by the distribution of stars. At smaller scales, the collisional nature of gas allows for the efficient formation of local gravitationally bound clumps as manifested in the $S$ parameter derived here. In this regard, while the formation process of the two features (clumps and large-scale) are likely different, the correlation seen in figure~\ref{fig:A1_S} suggests a causal link between the two processes. The ubiquity of star clusters (e.g.,~\cite{He2022}) and gas clumps (e.g.,~\cite{Zaragoza2014}) in interacting and colliding galaxies, whose large-scale stellar dynamics are susceptible to temporal and spatially varying gravitational potential, supports this assertion.
This picture is consistent with the results of recent simulations (e.g.,~\cite{Pettitt2018b}).

\subsection{Relationship between the galactic bar and gas concentration}
We presented in section~3.2 that barred spiral galaxies have higher central concentration than non-barred galaxies.  This result is consistent with past observational  (e.g.,~\cite{Sakamoto1999};~\cite{Sheth2005};~\cite{Kuno2007}) and theoretical (e.g.,~\cite{Wada1992};~\cite{Wada1995};~\cite{Regan2004};~\cite{Baba2022}) works that examine the impact of the galactic bar on the distribution and dynamics of gas.
 \citet{Wada1992} suggest that the gravitational influence of a galactic bar removes the angular momentum of molecular gas, leading to its accumulation and subsequent formation of an elongated ring structure. The self-gravity inherent in the gas further augments the moment of inertia within the accumulated structure, resulting in a rapid reduction of angular momentum. The latter process depends on the strength of the galactic bar.
\citet{Regan2004} demonstrated that bars of weaker strength exhibit negligible mass inflow into the inner kilo-parsec, signifying that weak bars have a minimal impact on the evolution of galaxies.  Conversely, stronger bars give rise to the formation of nuclear rings and induce substantial gas migration into the central kilo-parsec, thereby exerting a more pronounced influence on the evolution of the nuclear region. In a more recent investigation, \citet{Baba2022} utilized N-body/hydrodynamical simulations to illustrate the evolution and physical mechanisms underlying gas concentration at the center and the subsequent formation of the galactic bulge.

The results presented in this study reconfirm the close correlation between the presence of a bar and the concentration of nuclear gas. Nevertheless, within barred spiral galaxies, there exists a considerable variation of $C$ values, ranging from $3.0$ to $7.7$. Notably, $25\%$ of barred spiral galaxies exhibit $C$ values equal to or lower than the average $C$ value observed in non-barred spiral galaxies. 

To explore the potential influence of bar length on $C$ and the resultant variability in $C$ values, we plot the bar length ($R_\mathrm{bar}$) normalized by the isophotal radius $R_{25}$ ($R_\mathrm{bar}/R_{25}$)  against the corresponding $C$ values in figure~\ref{fig:C-BL4}. $R_\mathrm{bar}/R_{25}$ is calculated from the data in table 1 of \cite{Lang2020}.
The Spearman's rank correlation coefficient computed for this analysis is 0.63, with a $p$-value of $3.8\times 10^{-5}$. This correlation suggests that an extended bar structure is associated with higher $C$ values since a longer bar structure can collect molecular gas from a relatively large area, resulting in more efficient angular momentum transfer. This finding is qualitatively consistent with the outcomes of \citet{Kuno2007} who demonstrated that the bar structure efficiently facilitates the transport of molecular gas, while playing a minor role in drawing gas from outside the bar. 

Finally, we note two barred galaxies (NGC~4424 and NGC~4731) that do not follow the correlation presented in figure~\ref{fig:C-BL4}, most likely due to a recent interaction with another galaxy. NGC~4424 is a member of the Virgo Cluster (VCC;~\cite{Binggeli1985}). NGC~4731 belongs to the NGC~4697 group in the Virgo Southern Extension (\cite{Karachentsev2013}). The H$\alpha$ image of NGC~4424 shows a $\sim10~\mathrm{kpc}$ tail extending from the galactic center, a strong evidence of galaxy merger that had occurred  $\lesssim\mathrm{500Myr}$, and this galaxy is likely undergoing a ram pressure stripping event (\cite{Boselli2018}). NGC~4731 shows strong traces of interaction with the nearby elliptical galaxy NGC 4697 (e.g.,~\cite{Martin1997}).
Those two galaxies are added to the $C$ vs $R_\mathrm{bar}/R_{25}$ plot and presented in figure~\ref{fig:C-BL_VC}. Despite possessing elevated $A$ values (0.68 and 0.70 for NGC~4424 and NGC~4731, respectively), indicative of 
 possible influence of the interaction, 
these galaxies exhibit $C$ values that are significantly smaller, deviating from the trend. 
Indeed, CO~(\textit{J}\ =\ 2--1) is nearly absent in the center of NGC~4424, 
which may indicate the destruction or expulsion of molecular gas during the interaction. An alternate explanation for the observed deviation from the general trend is the delayed central concentration of gas. Such a delay may arise from the recent formation of the bar structure, as illustrated in a recent simulation by \citet{Iles2022}.

Figure~\ref{fig:C-BL_VC} shows the correlation between the Concentration ($C$) index and bar length for Virgo Cluster galaxies and non-Virgo galaxies. While overall there is no significant difference in the Virgo cluster and non-Virgo population as a whole, it is evident that some Virgo Cluster galaxies, including NGC 4424 which is possibly experiencing ram pressure stripping \citep{Boselli2018}, deviate downward from the main part of this correlation.  It is unclear why these galaxies display particularly low $C$ values, but one possibility is that the may have suffered from gas stripping by ram pressure or tidal interactions within the Virgo Cluster.

\subsection{Qualitative comparison with \citet{Davis2022} correlation measurements}
Here we attempt to quantitatively compare the results of the correlation measurement of \citet{Davis2022} to this study. However, we note that the sample galaxies used in both analyses are different; the analysis of \citet{Davis2022} is based on the sample of 60 PHANGS galaxies and 26 WISDOM galaxies, which includes 15 ETGs (elliptical and lenticular galaxies) and significantly larger than the four ETGs (lenticular galaxies only) used in our analysis. In the case of ETGs, both $A$ and $S$ tend to be low, which is a critical factor in the correlation measurement. Table~\ref{tab:004} shows correlation measurements between each CAS morphological parameter and correlation measurements between each CAS and physical parameter.

The definition of the index in~\citet{Davis2022} called "Smoothness" is the same as our Clumpiness index. Our measurements generally follow the same trend as those reported by~\citet{Davis2022}. However, three specific comparisons show different trends: $\log_{10}{M_*}$ vs. $A$, $\log_{10}({M_\mathrm{H_2}/M_{*}})$ vs. $A$, and $\log_{10}({M_\mathrm{H_2}/M_{*}})$ vs. $S$. ~\citet{Davis2022} reported a Spearman's rank correlation coefficient of -0.54 for $\log_{10}{M_*}$ vs. $A$. The corresponding $p$-value is $1 \times 10^{-7}$. Nonetheless, our study did not find a significant correlation between the two (a Spearman's rank correlation coefficient of $-0.21$, with a $p$-value of $0.07 (>0.05)$). The scatter plot in Figure 3 of ~\citet{Davis2022} shows that the WISDOM ETGs occupy a region characterized by low $A$ values and high stellar mass, which may explain the difference between the correlation coefficients reported by~\citet{Davis2022} and our study. Regarding the relationship between $\log_{10}({M_\mathrm{H_2}/M_{*}})$ and $A$, ~\citet{Davis2022} reported a Spearman's rank correlation coefficient of 0.31 with a $p$-value of 0.004, but our measurements show no significant correlation. The scatter plot in Figure 4 of~\citet{Davis2022} indicates that the WISDOM ETGs are located in a region with low $A$ values and low molecular gas mass. This difference in the correlation measurements is likely due to the specific characteristics of these ETGs. Interestingly, for the correlation between $\log_{10}({M_\mathrm{H_2}/M_{*}})$ and $S$,~\citet{Davis2022} reported a Spearman's rank correlation coefficient of -0.03 with a $p$-value of $0.8 (>0.05)$, whereas our measurement yielded a Spearman's rank correlation coefficient of -0.42 with a $p$-value of $2\times 10^{-4}$, indicating a significant negative correlation. Figure 4 of~\citet{Davis2022} suggests that the correlation between $\log_{10}({M_\mathrm{H_2}/M_{*}})$ and $S$ is not significant because the distribution of WISDOM ETGs in terms of $\log_{10}({M_\mathrm{H_2}/M_{*}})$ and $S$ values is low. In summary, our measurements generally exhibit a qualitatively similar trend to the results of~\citet{Davis2022}. However, some differences are observed, and they may arise due to the morphological and physical characteristics of the WISDOM ETGs.

\subsection{Gas Morphological Evolution in the CAS plane}
\begin{figure}
 \begin{center}
  \includegraphics[width=80mm,height=80mm]{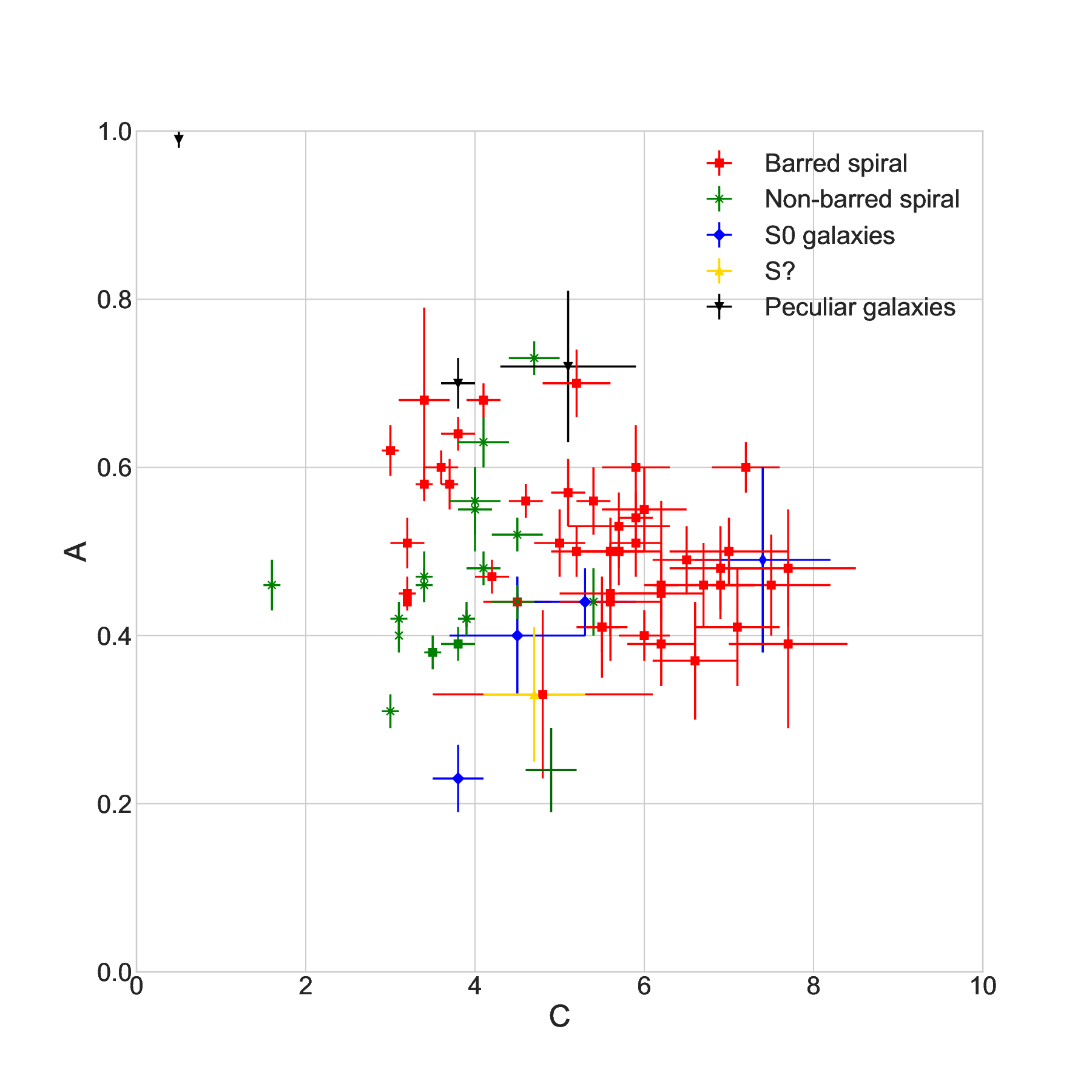}
 \end{center}
 \caption{
 Scatter plot of the relationship between $C$ and $A$ on horizontal and vertical axes. Barred spiral galaxies are shown as red squares, non-barred spiral galaxies as green crosses, S0 (lenticular) galaxies as blue diamonds, an S? galaxy as a gold triangle, and peculiar types as black inverted triangles.}
 \label{fig:A1_C}
\end{figure}

\begin{figure}
 \begin{center}
  \includegraphics[width=80mm,height=80mm]{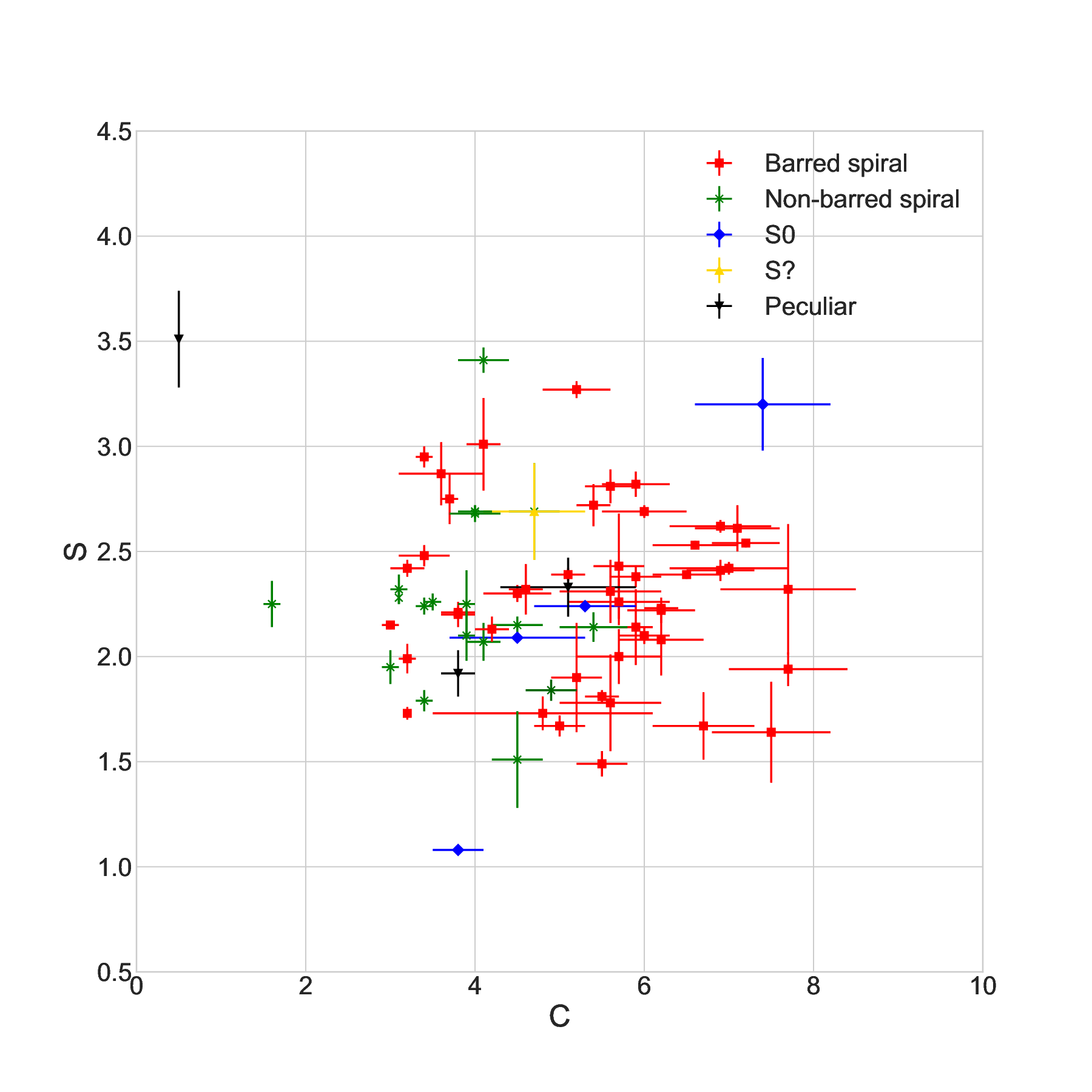}
 \end{center}
 \caption{
Scatter plot of the relationship between $C$ and $S$ on horizontal and vertical axes. The symbols are the same as in figure~\ref{fig:A1_C}.}
 \label{fig:S_C}
\end{figure}

\begin{figure}
 \begin{center}
  \includegraphics[width=80mm,height=80mm]{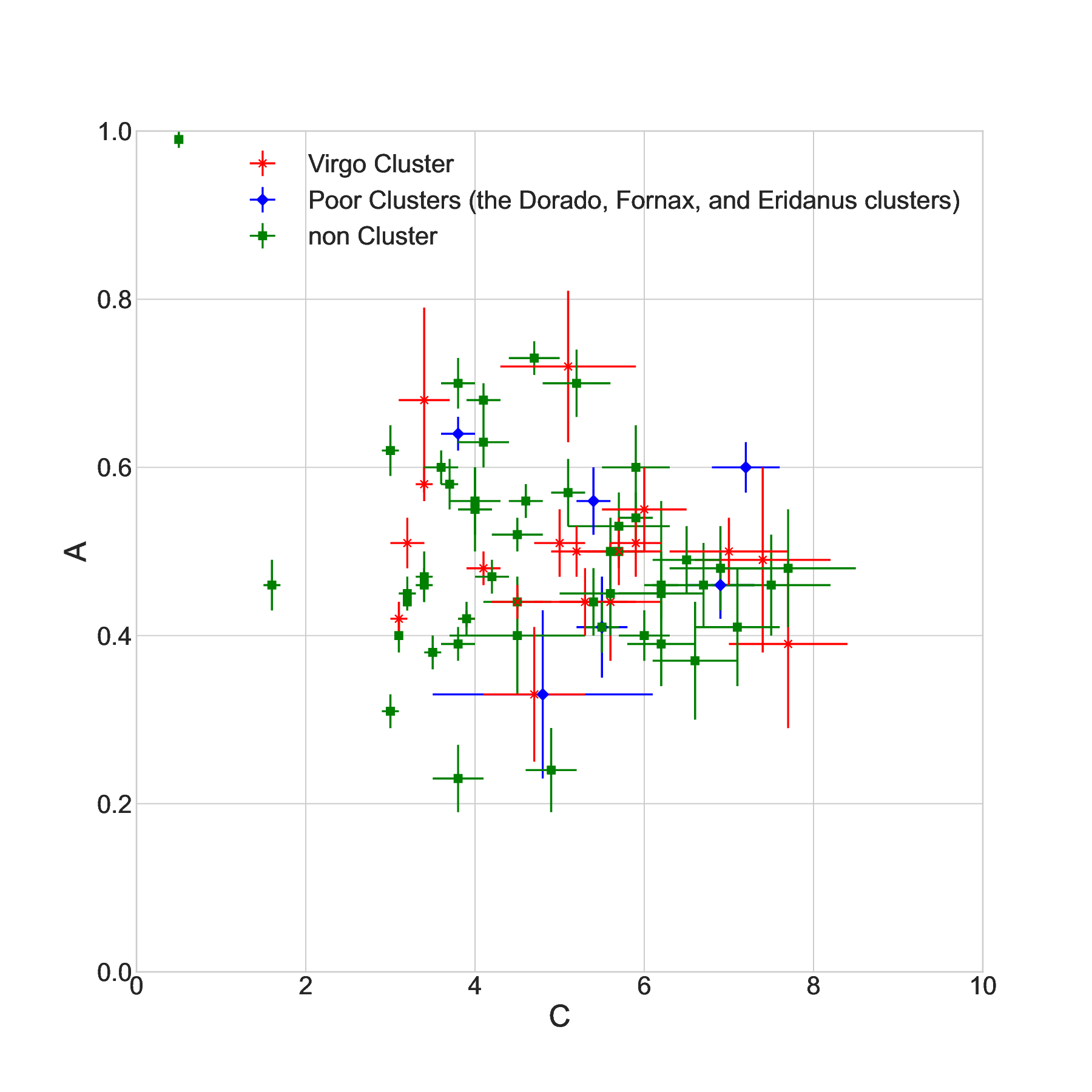}
 \end{center}
 \caption{
 Scatter plot of the relationship between $C$ and $A$ on horizontal and vertical axes, comparison between Virgo cluster galaxies, poor cluster galaxies and the other galaxies. The Virgo Cluster galaxies are shown as red crosses, poor cluster galaxies as blue diamonds, non cluster galaxies as green squares.}
 \label{fig:A1_C_VC}
\end{figure}

\begin{figure}
 \begin{center}
  \includegraphics[width=80mm,height=80mm]{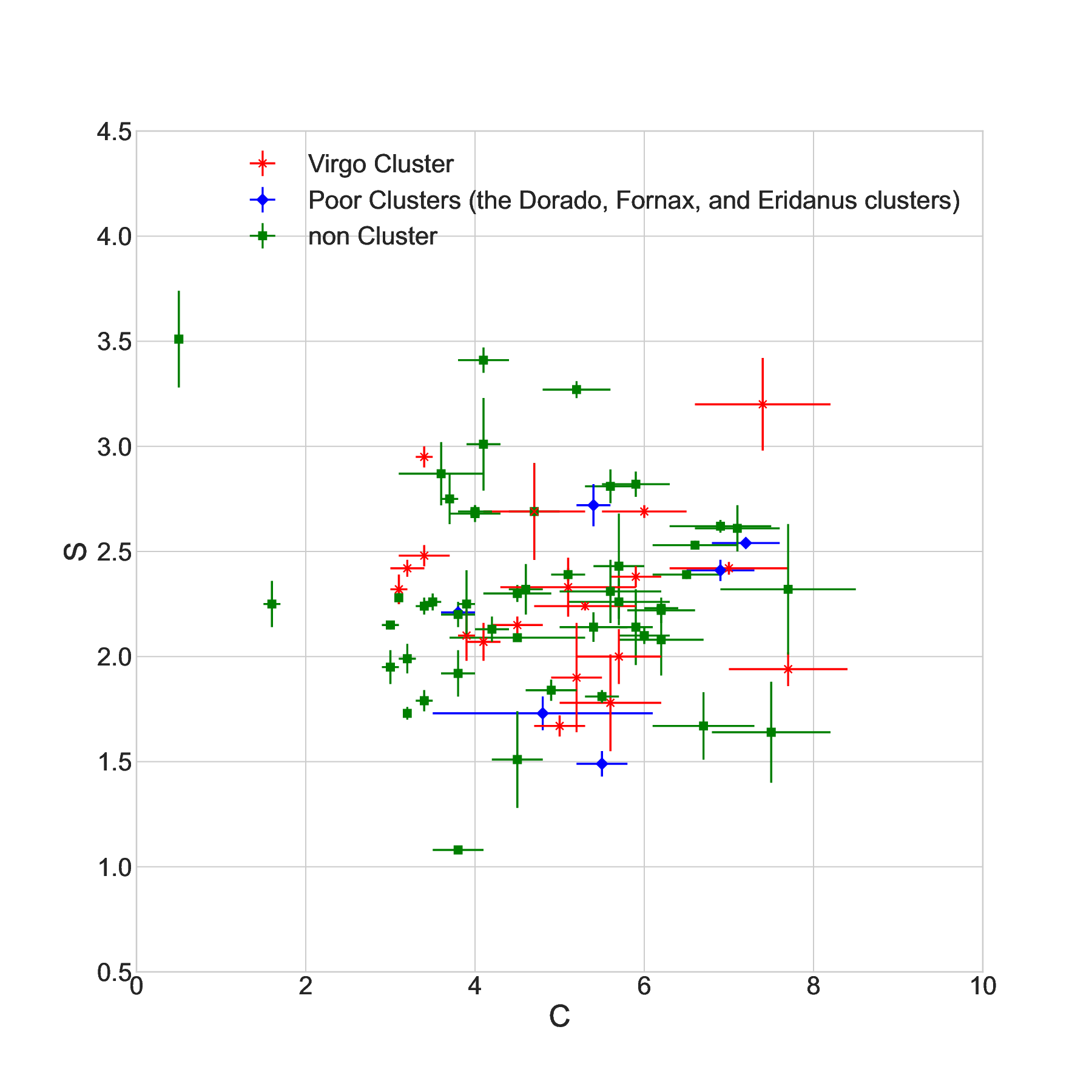}
 \end{center}
 \caption{
Scatter plot of the relationship between $C$ and $S$ on horizontal and vertical axes, comparison between Virgo cluster galaxies, poor cluster galaxies and the other galaxies. The symbols are the same as in figure~\ref{fig:A1_C_VC}.}
 \label{fig:S_C_VC}
\end{figure}

\begin{figure*}
 \begin{center}
  \includegraphics[height=60mm]{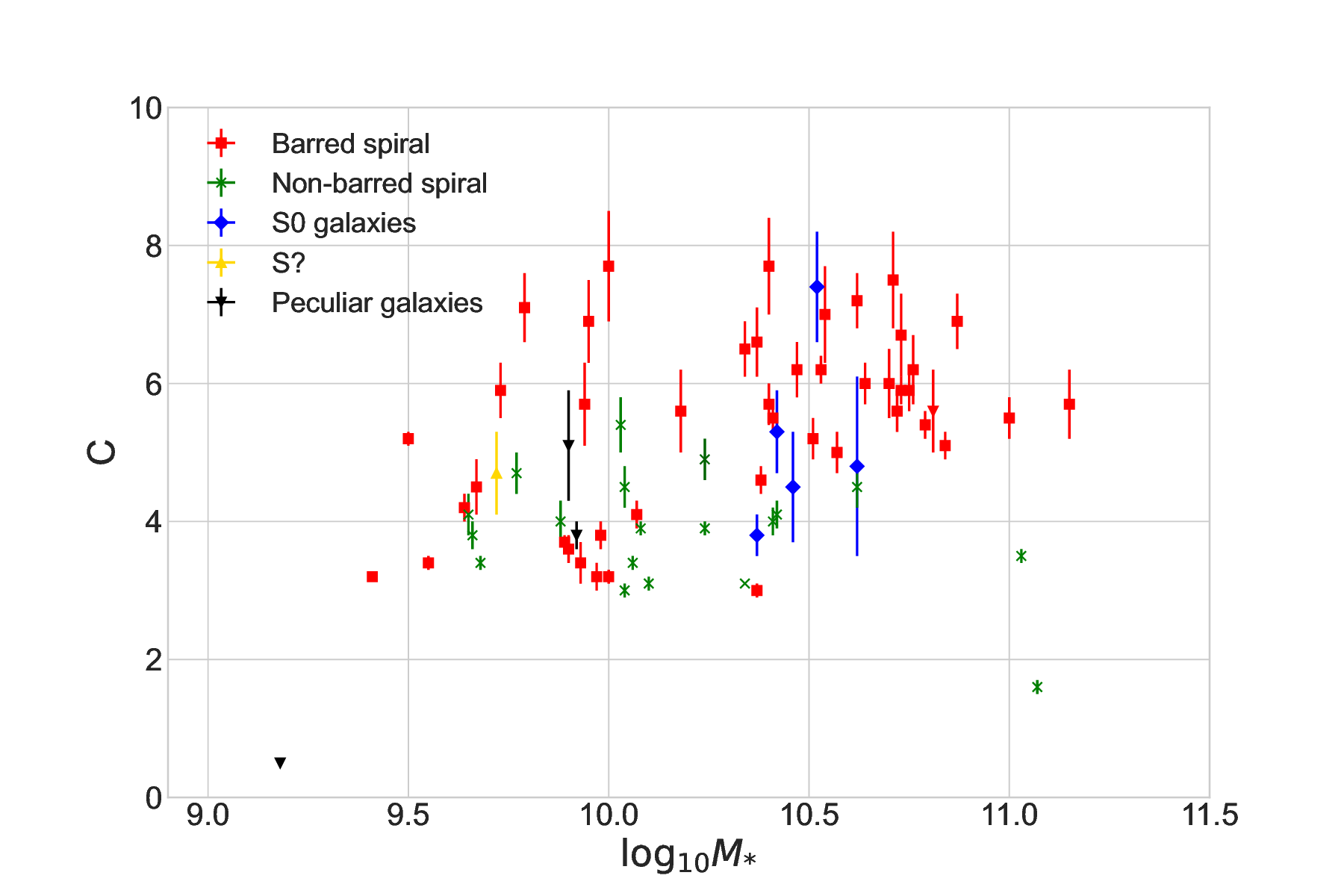}
 \end{center}
 \caption{Scatter plot of the relationship between stellar mass ($\log_{10}M_{*}$) and $C$ on horizontal and vertical axes. Barred spiral galaxies are shown as red squares, non-barred spiral galaxies as green crosses, S0 (lenticular) galaxies as blue diamonds, an S? galaxy as a gold triangle, and peculiar types as black inverted triangles.}
 \label{fig:M_st vs. C}
\end{figure*}

\begin{figure*}
 \begin{center}
  \includegraphics[height=60mm]{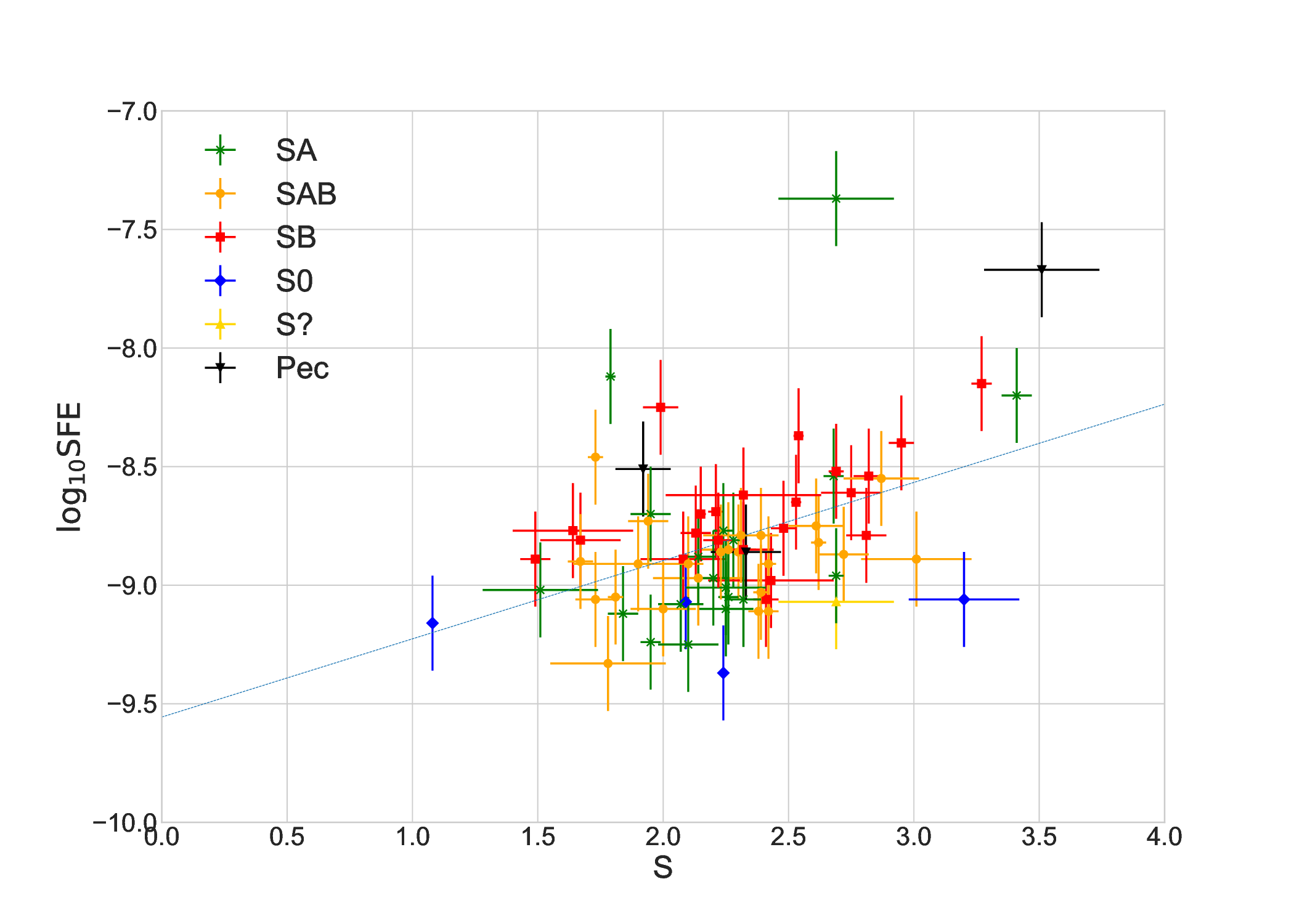}
 \end{center}
 \caption{
 Scatter plot of the relationship between $S$ and $\log_{10}\mathrm{SFE}$ on horizontal and vertical axes. In this figure, barred spiral galaxies are shown as red squares, non-barred spiral galaxies as green crosses, S0 (lenticular) galaxies as blue diamonds, a S? galaxy as gold triangle and peculiar types as black inverted triangles.}
 \label{fig:SFE_S}
\end{figure*}

\begin{figure*}
 \begin{center}
  \includegraphics[height=68mm]{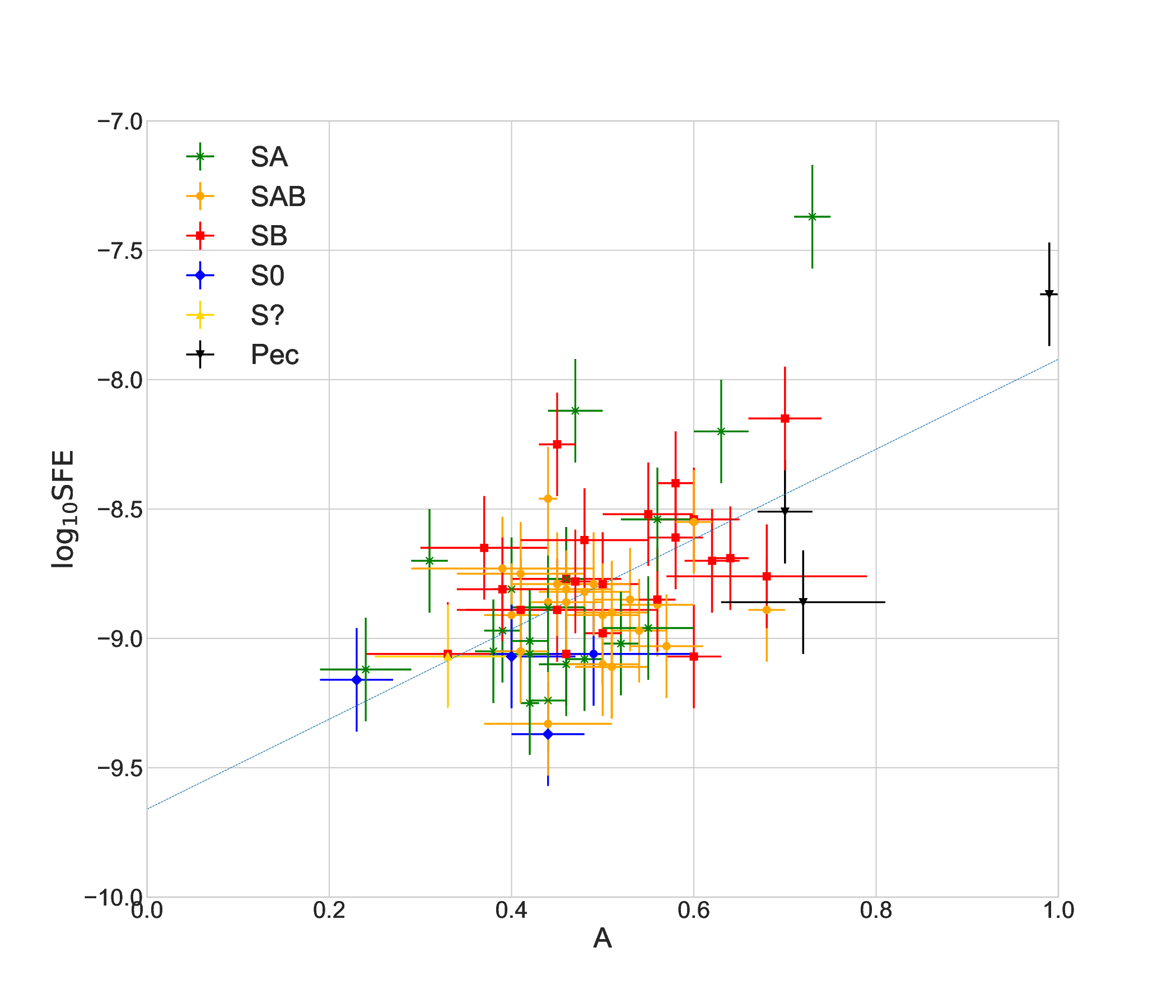}
 \end{center}
 \caption{
 Scatter plot of the relationship between $A$ and $\log_{10}\mathrm{SFE}$ on horizontal and vertical axes. The symbols are the same as in figure~\ref{fig:SFE_S}. The sloping thin dotted line is the regression line by the least-squares method.}
 \label{fig:SFE_A}
\end{figure*}

\begin{figure*}
 \begin{center}
  \includegraphics[width=140mm,height=100mm]{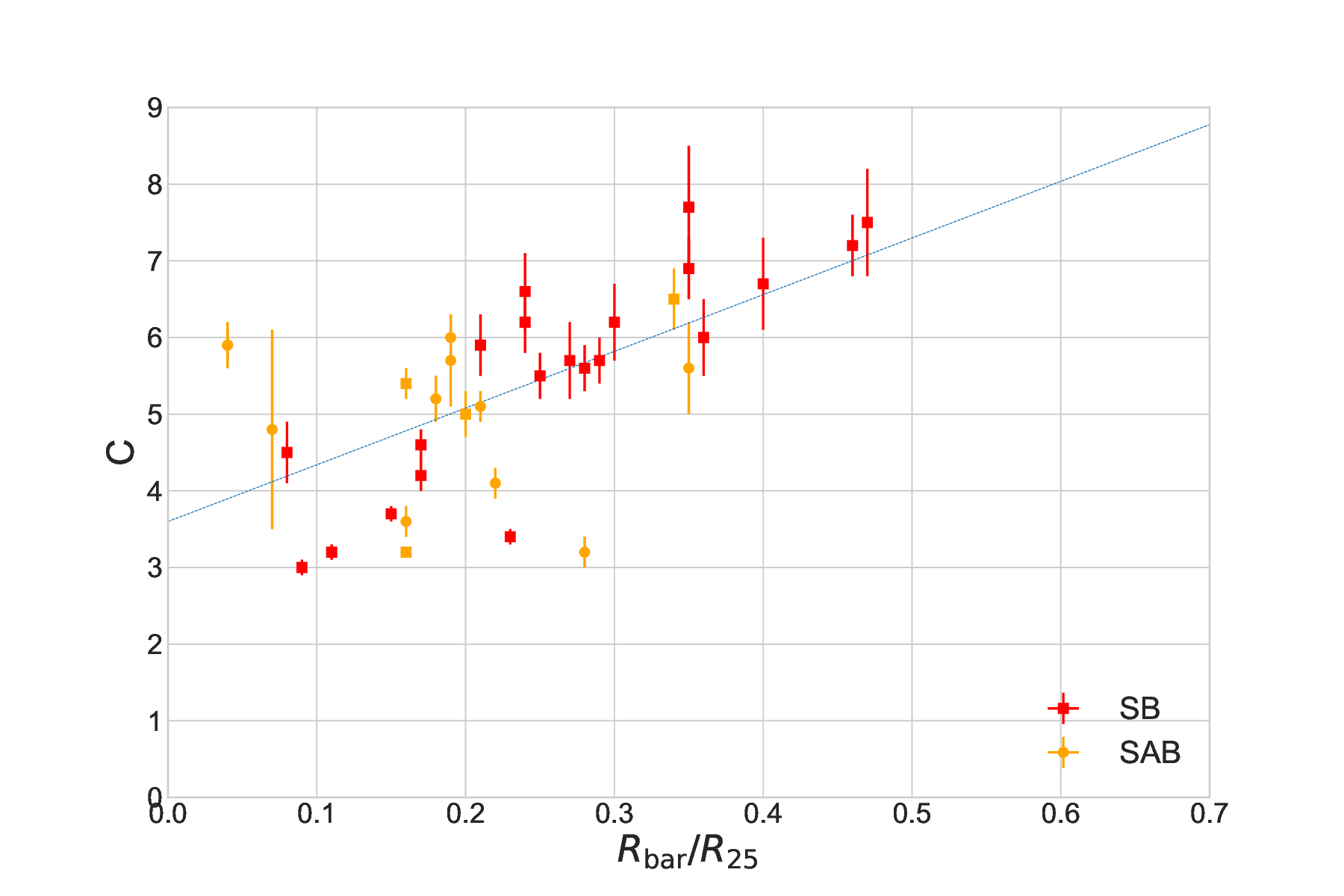}
 \end{center}
 \caption{Bar length ($R_\mathrm{bar}/R_{25}$) vs. Concentration. SB galaxies are shown as red squares, SAB galaxies as orange circles. The sloping thin dotted line is the regression line by the least-squares method.}

 \label{fig:C-BL4}
\end{figure*}

\begin{figure*}
 \begin{center}
  \includegraphics[width=140mm,height=100mm]{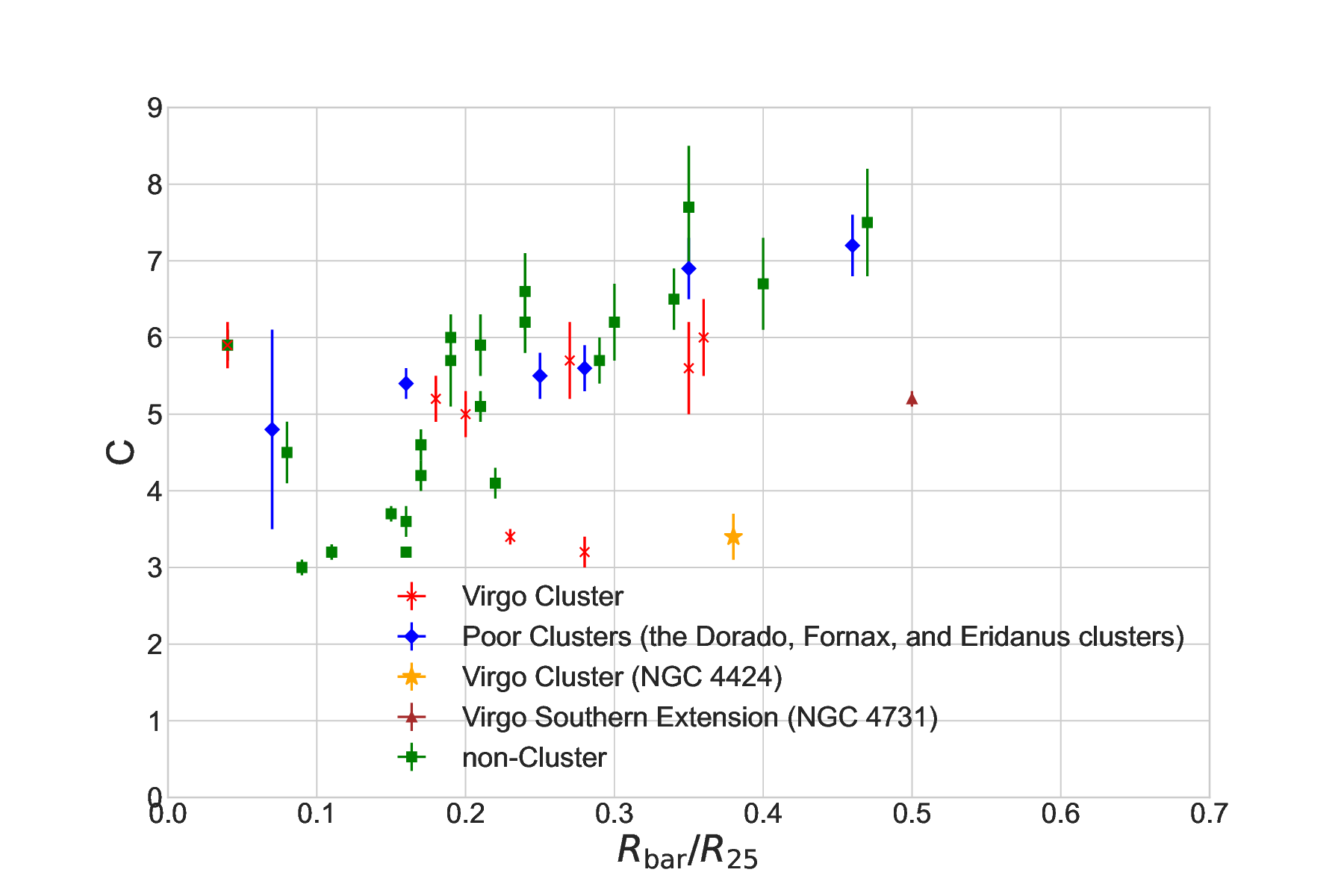}
 \end{center}
 \caption{Bar length ($R_\mathrm{bar}/R_{25}$) vs. Concentration by galaxies' environments. The Virgo Cluster galaxies are represented by red crosses. Among them, NGC 4424 (a galaxy currently undergoing ram-pressure stripping) is marked with a yellow star, and NGC 4731 (a galaxy in the Virgo Southern Extension, displaying strong signs of interaction) is marked with a brown triangle. Poor clusters are shown as blue diamonds, Non-Virgo Cluster galaxies are shown as green squares. }

 \label{fig:C-BL_VC}
\end{figure*}

\begin{figure*}
 \begin{center}
  \includegraphics[width=140mm,height=80mm]{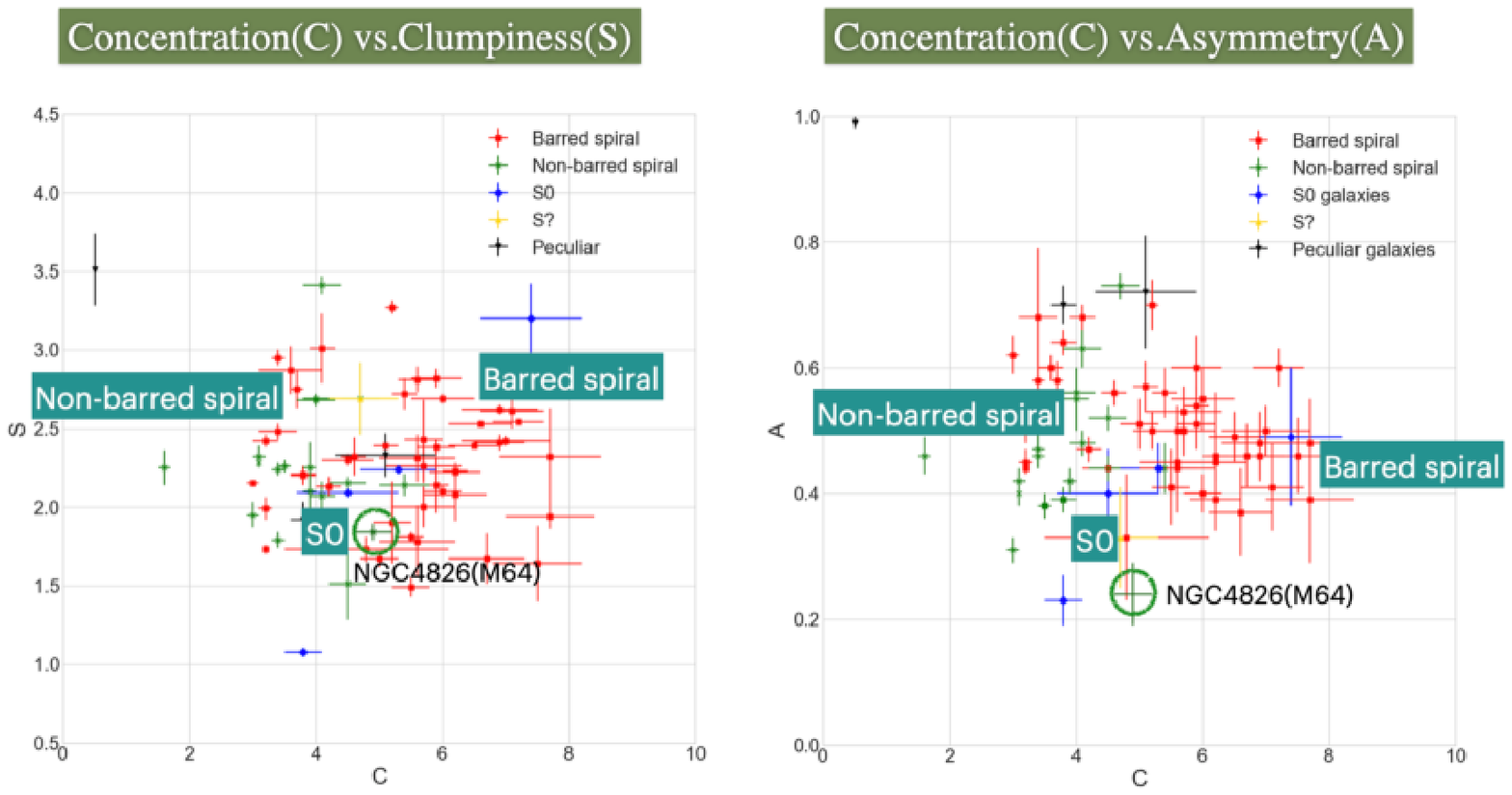}
 \end{center}
 \caption{
The scatter plot of $C$ versus $S$ (left) and $C$ versus $A$ (right). Barred spiral galaxies are shown as red squares, non-barred spiral galaxies as green crosses, S0 (lenticular) galaxies as blue diamonds, an S? galaxy as a gold triangle, and peculiar types as black inverted triangles. Barred spiral galaxies, non-barred spiral galaxies, and S0 galaxies occupy distinct regions in the CAS plane, possibly related to the evolution of molecular gas distribution within each galaxy. We note NGC 4826, which is a galaxy in the process of merging and transforming into an S0 galaxy, is located close to the general area in the CAS plane where S0 galaxies occupy.
}
 \label{fig:Concentration.3}
\end{figure*}

Close inspection of the scatter plots showing the relationships between $C$-$A$ and $C$-$S$ presented in figure~\ref{fig:A1_C} and ~\ref{fig:S_C} suggest noticeable patterns -- non-barred spiral galaxies (SA), barred spiral galaxies (SB + SAB) and lenticular galaxies (S0) generally occupy distinct regions in the CAS  space.  In the past, the CAS parameters derived by optical/IR images were used for galaxy classification (e.g., ~\cite{Bershady2000};~\cite{Conselice2003}), but the results in figure~\ref{fig:A1_C} and~\ref{fig:S_C} emphasizes the potential utility of these parameters in effectively distinguishing and characterizing galaxy types based on their molecular gas structure and features. The morphological evolution of molecular gas along the CAS parameter space is shown in figure~\ref{fig:Concentration.3}.

Bar structures are thought to arise from either bar instabilities \citep{Sheth2008} or gravitational tidal interactions \citep{Lokas2016, Gajda2018}, implying that the evolution of galaxies often involves a transition from an SA to SB + SAB. Similarly, S0 galaxies, characterized by the absence of spiral arms and a dominance of galactic rotation in their kinematics, are believed to originate from mergers between equal mass galaxies \citep{Querejeta2015}. As an example in our sample, NGC~4826 is classified as an SA galaxy but it is, in fact, believed to be transforming from SA to S0, based on HI kinematics \citep{Braun1992} and the characteristics of the stellar population at the outermost edge of the galaxy disk \citep{Watkins2016}. NGC~4826 is relatively isolated but belongs to the loose Canes Venatici (CVn) I group (\cite{Israel2009}).
NGC~4826 occupies a region in figure~\ref{fig:A1_C},~\ref{fig:S_C} and~\ref{fig:Concentration.3} that generally aligns with S0 galaxies, providing additional observational support for its transitional nature from SA to S0.

\begin{figure*}
 \begin{center}
  \includegraphics[width=160mm]{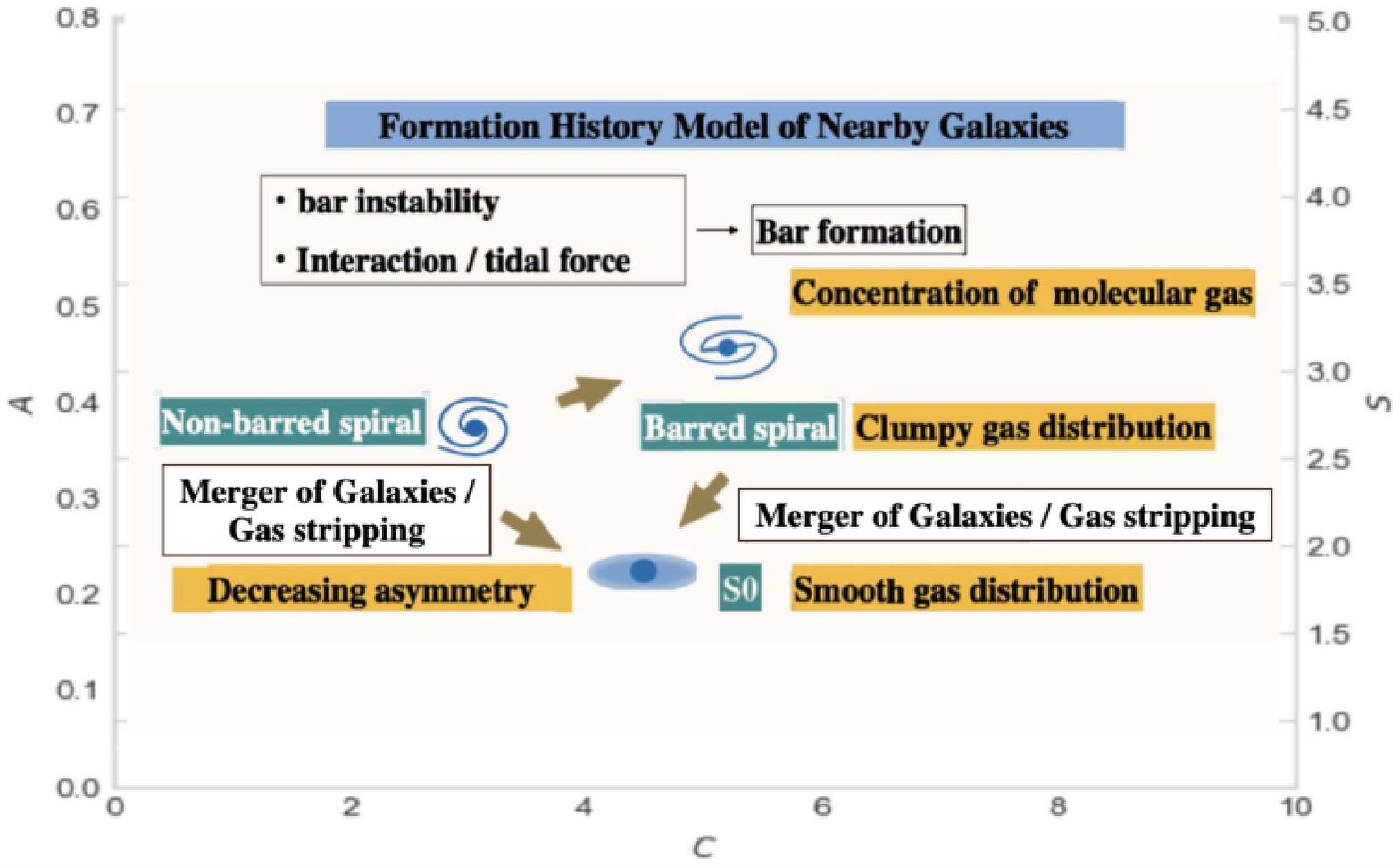}
 \end{center}
 \caption{Formation and evolution history model of nearby star-forming galaxies based on molecular gas morphology. The galaxies of each type (Barred spiral, Non-barred spiral and S0) are roughly displayed in CAS plane. The horizontal axis is $C$, the left vertical axis is $A$, and the right vertical axis is $S$.
}
 \label{fig:Formation_History_Model}
\end{figure*}

The evolutionary history of nearby star-forming galaxies is thought to be primarily driven by bar formation, galaxy mergers and gas stripping, a picture that seems plausible in the characteristics of the molecular gas distribution as indicated by the location of the different types of galaxies plotted in the CAS space (Figure~\ref{fig:Formation_History_Model}). S0 galaxies are more abundant than spiral galaxies in high-density regions of galaxy clusters (e.g.,~\cite{Dressler1980};~\cite{Goto2003}). The high-density environment in a galaxy cluster has a high probability of close encounters and collisions between galaxies, which can be the reason for the high number of S0 galaxies. However, the formation of S0 galaxies remains uncertain. In present-day rich clusters, mergers are less likely to occur due to the high velocities of satellite galaxies around the cluster center (e.g.,~\cite{Binggeli1993}). Interactions in these environments tend to result in high-speed encounters (e.g.,~\citet{Binney2008}). Recently, gas stripping, in addition to mergers, has been proposed as a potential formation pathway for S0 galaxies (e.g.,~\citet{Coccato2020};~\cite{Deeley2020};~\cite{Deeley2021}). Gas stripping may play a significant role in the formation of S0 galaxies in rich clusters.
This general picture, however, may not apply to all galaxies in the field where the number density of galaxies is significantly lower than that in clusters. A thorough morphological survey across the history of the universe is necessary to address how the environment affects the morphological transformation of galaxies in general. High-resolution cosmological simulations may also offer some insights into this scenario.

\section{Conclusion}

We present a quantitative and statistical analysis of the molecular gas morphology in 73 nearby galaxies using high spatial resolution CO~(\textit{J}\ =\ 2--1) data obtained from the Atacama Large Millimeter/submillimeter Array (ALMA) by the PHANGS large program. We utilize the model-independent, non-parametric CAS analysis often used in characterizing optical galaxy morphology. The main points are summarized below.
 
1. We find a correlation between Clumpiness ($S$) and Asymmetry ($A$), with a Spearman’s rank correlation coefficient of 0.52 and the associated $p$-value of $2\times 10^{-6}$, suggesting that galaxies with significant biases or distortions in the molecular gas distribution are susceptible to local clump formation. While the formation process of the two features (clumps and large-scale) are likely different, this moderate correlation suggests a causal link between the two processes. 

2. We find that the Concentration ($C$) of barred galaxies is significantly higher than non-barred galaxies, suggesting an overall trend that barred spiral galaxies can facilitate gas inflow to the center, as suggested by previous studies. Furthermore, we find a positive correlation between $C$ and the bar length ($R_\mathrm{bar}/R_{25}$), with the Spearman’s rank correlation coefficient of 0.63 and a corresponding $p$-value of $3.8\times 10^{-5}$. This suggests that a longer bar structure can collect molecular gas from a relatively large area, resulting in more efficient angular momentum exchange. 

3. The large sample size studied here allows us to depict the evolutionary history of molecular gas distribution in the CAS plane. In the past, the  CAS parameters derived by optical/IR images were used for galaxy classification, but the results presented in this study emphasize the potential utility of these parameters in effectively distinguishing and characterizing galaxy types based on their molecular gas structure and features. 
 
Finally, we advocate that the quantitative and statistical analysis of the distribution morphology of molecular gas using non-parametric indices acquired in this study can be applied to the morphological classification of spatially resolved images of gas and dust in distant galaxies. These data are now available using ALMA and deferred to a future study.

\begin{table*}
    \centering
    \tbl{Concentration, Asymmetry and Clumpiness. The associated errors and Galaxy types are shown.}{%
    \begin{tabular}{lccccccc}
   galaxy id & Concentration ($C$) & Asymmetry ($A$) & Clumpiness ($S$) & $\Delta C$ & $\Delta A$ & $\Delta S$ & Galaxy Type $*$ \\ \hline
IC 1954 & 4.5 & 0.44 & 2.30 & 0.4 & 0.01 & 0.04 & SB \\ 
IC 5273 & 5.9 & 0.60 & 2.82 & 0.4 & 0.05 & 0.06 & SB \\ 
IC 5332 & 3.4 & 0.47 & 1.79 & 0.1 & 0.03 & 0.02 & SA \\ 
NGC 628 & 3.1 & 0.40 & 2.28 & 0.0 & 0.02 & 0.03 & SA \\ 
NGC 685 & 4.1 & 0.68 & 3.01 & 0.2 & 0.02 & 0.22 & SAB \\ 
NGC 1087 & 5.7 & 0.53 & 2.26 & 0.6 & 0.04 & 0.11 & SAB \\ 
NGC 1097 & 6.2 & 0.45 & 2.08 & 0.5 & 0.11 & 0.17 & SB \\ 
NGC 1300 & 7.2 & 0.60 & 2.54 & 0.4 & 0.03 & 0.02 & SB \\ 
NGC 1317 & 4.8 & 0.33 & 1.73 & 0.6 & 0.10 & 0.08 & SAB \\ 
NGC 1365 & 5.5 & 0.41 & 1.49 & 0.3 & 0.06 & 0.06 & SB \\ 
NGC 1385 & 3.8 & 0.64 & 2.21 & 0.2 & 0.02 & 0.03 & SB \\ 
NGC 1433 & 6.9 & 0.46 & 2.41 & 0.4 & 0.04 & 0.05 & SB \\ 
NGC 1511 & 3.8 & 0.70 & 1.92 & 0.2 & 0.03 & 0.11 & pec \\
NGC 1512 & 5.6 & 0.50 & 2.81 & 0.3 & 0.04 & 0.08 & SB \\ 
NGC 1546 & 3.8 & 0.23 & 1.08 & 0.3 & 0.04 & 0.02 & S0 \\ 
NGC 1559 & 3.0 & 0.62 & 2.15 & 0.1 & 0.03 & 0.02 & SB \\ 
NGC 1566 & 5.4 & 0.56 & 2.72 & 0.2 & 0.04 & 0.10 & SAB \\ 
NGC 1637 & 6.9 & 0.48 & 2.62 & 0.6 & 0.05 & 0.03 & SAB \\ 
NGC 1672 & 6.7 & 0.46 & 1.67 & 0.6 & 0.05 & 0.16 & SB \\ 
NGC 1792 & 4.5 & 0.52 & 1.51 & 0.3 & 0.02 & 0.23 & SA \\ 
NGC 1809 & 4.7 & 0.73 & 2.69 & 0.3 & 0.02 & 0.23 & SA \\ 
NGC 2090 & 3.0 & 0.31 & 1.95 & 0.1 & 0.02 & 0.08 & SA \\ 
NGC 2283 & 3.7 & 0.58 & 2.75 & 0.1 & 0.03 & 0.12 & SB \\ 
NGC 2566 & 7.5 & 0.46 & 1.64 & 0.7 & 0.06 & 0.24 & SB \\ 
NGC 2775 & 1.6 & 0.46 & 2.25 & 0.1 & 0.03 & 0.11 & SA \\ 
NGC 2835 & 3.2 & 0.45 & 1.99 & 0.1 & 0.02 & 0.07 & SB \\ 
NGC 2903 & 6.0 & 0.40 & 2.10 & 0.3 & 0.03 & 0.04 & SAB \\ 
NGC 2997 & 5.9 & 0.54 & 2.14 & 0.2 & 0.03 & 0.18 & SAB \\ 
NGC 3059 & 4.6 & 0.56 & 2.32 & 0.2 & 0.02 & 0.12 & SB \\ 
NGC 3137 & 4.0 & 0.56 & 2.68 & 0.3 & 0.04 & 0.04 & SA \\ 
NGC 3239 & 0.5 & 0.99 & 3.51 & 0.0 & 0.01 & 0.23 & pec \\
NGC 3351 & 6.6 & 0.37 & 2.53 & 0.5 & 0.07 & 0.02 & SB \\ 
NGC 3507 & 5.7 & 0.50 & 2.43 & 0.3 & 0.02 & 0.25 & SB \\ 
NGC 3511 & 5.4 & 0.44 & 2.14 & 0.4 & 0.04 & 0.07 & SA \\ 
NGC 3521 & 3.5 & 0.38 & 2.26 & 0.1 & 0.02 & 0.04 & SA \\ 
NGC 3596 & 3.8 & 0.39 & 2.20 & 0.2 & 0.02 & 0.06 & SA \\ 
NGC 3621 & 3.4 & 0.46 & 2.24 & 0.1 & 0.02 & 0.04 & SA \\ 
NGC 3626 & 4.5 & 0.40 & 2.09 & 0.8 & 0.07 & 0.02 & S0 \\ 
NGC 3627 & 5.1 & 0.57 & 2.39 & 0.2 & 0.04 & 0.03 & SAB \\ 
NGC 4207 & 4.7 & 0.33 & 2.69 & 0.6 & 0.08 & 0.23 & S? \\ 
NGC 4254 & 4.1 & 0.48 & 2.07 & 0.2 & 0.02 & 0.09 & SA \\ 
NGC 4293 & 7.4 & 0.49 & 3.20 & 0.8 & 0.11 & 0.22 & S0 \\ 
NGC 4298 & 4.5 & 0.44 & 1.95 & 0.3 & 0.02 & 0.04 & SA \\ 
NGC 4303 & 5.2 & 0.50 & 1.90 & 0.3 & 0.03 & 0.26 & SAB \\ 
NGC 4321 & 5.9 & 0.51 & 2.38 & 0.3 & 0.04 & 0.04 & SAB \\ 
NGC 4424 & 3.4 & 0.68 & 2.48 & 0.3 & 0.11 & 0.05 & SB \\ 
NGC 4457 & 5.3 & 0.44 & 2.24 & 0.6 & 0.04 & 0.02 & S0 \\ 
NGC 4496A & 3.4 & 0.58 & 2.95 & 0.1 & 0.02 & 0.05 & SB \\ 
NGC 4535 & 7.0 & 0.50 & 2.42 & 0.7 & 0.04 & 0.03 & SAB \\ \hline
    \end{tabular}}
    \label{tab:001}
    \begin{tabnote}
\hangindent6pt\noindent
\hbox to6pt{\footnotemark[$*$]\hss}\unskip%
 Galaxy Type used the following three papers as references. \citet{RC3}, \citet{Buta2015} and \citet{Stuber2023}.
    \end{tabnote}
\end{table*}

\begin{table*}
    \setcounter{table}{0}
    \centering
    \caption{(Continued)}
    \begin{tabular}{lccccccc}
   galaxy id & Concentration ($C$) & Asymmetry ($A$) & Clumpiness ($S$) & $\Delta C$ & $\Delta A$ & $\Delta S$ & Galaxy Type $*$\\ \hline
NGC 4536 & 7.7 & 0.39 & 1.94 & 0.7 & 0.10 & 0.08 & SAB \\ 
NGC 4540 & 3.2 & 0.51 & 2.42 & 0.2 & 0.04 & 0.04 & SAB \\ 
NGC 4548 & 6.0 & 0.55 & 2.69 & 0.5 & 0.05 & 0.03 & SB \\ 
NGC 4569 & 5.6 & 0.44 & 1.78 & 0.6 & 0.07 & 0.23 & SAB \\ 
NGC 4571 & 3.1 & 0.42 & 2.32 & 0.1 & 0.02 & 0.07 & SA \\ 
NGC 4579 & 5.7 & 0.50 & 2.00 & 0.5 & 0.04 & 0.13 & SAB \\ 
NGC 4654 & 5.0 & 0.51 & 1.67 & 0.3 & 0.04 & 0.04 & SAB \\ 
NGC 4689 & 3.9 & 0.42 & 2.10 & 0.1 & 0.01 & 0.12 & SA \\ 
NGC 4694 & 5.1 & 0.72 & 2.33 & 0.4 & 0.09 & 0.14 & pec \\
NGC 4731 & 5.2 & 0.70 & 3.27 & 0.4 & 0.04 & 0.04 & SB \\ 
NGC 4781 & 4.2 & 0.47 & 2.13 & 0.2 & 0.02 & 0.06 & SB \\ 
NGC 4826 & 4.9 & 0.24 & 1.84 & 0.3 & 0.05 & 0.06 & SA \\ 
NGC 4941 & 5.6 & 0.45 & 2.31 & 0.6 & 0.05 & 0.15 & SAB \\ 
NGC 4951 & 7.1 & 0.41 & 2.61 & 0.5 & 0.07 & 0.11 & SAB \\ 
NGC 5042 & 3.6 & 0.60 & 2.87 & 0.2 & 0.02 & 0.15 & SAB \\ 
NGC 5068 & 3.2 & 0.44 & 1.73 & 0.0 & 0.01 & 0.03 & SAB \\ 
NGC 5134 & 4.0 & 0.55 & 2.69 & 0.2 & 0.05 & 0.03 & SA \\ 
NGC 5236 & 6.2 & 0.46 & 2.23 & 0.2 & 0.04 & 0.02 & SAB \\ 
NGC 5248 & 5.5 & 0.41 & 1.81 & 0.2 & 0.03 & 0.03 & SAB \\ 
NGC 5530 & 3.9 & 0.42 & 2.25 & 0.1 & 0.02 & 0.16 & SA \\ 
NGC 5643 & 6.5 & 0.49 & 2.39 & 0.4 & 0.04 & 0.02 & SAB \\ 
NGC 6300 & 6.2 & 0.39 & 2.22 & 0.4 & 0.05 & 0.06 & SB \\ 
NGC 7456 & 4.1 & 0.63 & 3.41 & 0.3 & 0.03 & 0.06 & SA \\ 
NGC 7496 & 7.7 & 0.48 & 2.32 & 0.8 & 0.07 & 0.31 & SB \\ \hline
    \end{tabular}
    \begin{tabnote}
\hangindent6pt\noindent
\hbox to6pt{\footnotemark[$*$]\hss}\unskip%
 Galaxy Type used the following three papers as references. \citet{RC3}, \citet{Buta2015} and \citet{Stuber2023}.
    \end{tabnote}

\end{table*}

\begin{table*}
    \centering
    \tbl{Statistical results of the CAS parameters.\footnotemark[$*$]}{%
    \begin{tabular}{cccccccccc}
Mean $C$ & Mean $A$ & Mean $S$ & $\sigma_C$ & $\sigma_A$ & $\sigma_S$ & Median $C$ & Median $A$ & Median $S$ \\ \hline
4.9 & 0.50 & 2.29 & 1.5 & 0.12 & 0.45 & 4.9 & 0.48 & 2.26 \\ \hline
    \end{tabular}}
    \label{tab:002}
    \begin{tabnote}
\hangindent6pt\noindent
\hbox to6pt{\footnotemark[$*$]\hss}\unskip%
 Means, standard deviations, and medians for each of the CAS indicators. $\sigma$ is the standard deviation value.
    \end{tabnote}
\end{table*}

\begin{table*}[t]
    \centering
    \tbl{PHANGS-ALMA galaxy properties.\footnotemark[$*$]}{%
    \begin{tabular}{p{18mm}ccccccc}
galaxy id & $\log_{10}({M_\mathrm{H_2}/M_*})$ & $\log_{10}({M_{*}}/{M_\odot})$ & $\log_{10}\mathrm{sSFR}(\mathrm{yr^{-1}})$ & $\log_{10}\mathrm{SFE\mathrm({yr^{-1}}})$ & $\log_{10}({M_\mathrm{H_I}}/{M_{\odot}})$ & $d$ (Mpc)\\
 &(1)&(2)&(3)&(4)&(5)&(6) \\
\hline
IC 1954 & -1.25 & 9.67 & -10.11 & -8.86 & 8.85 & $12.80\ \pm\ 2.17$\\
IC 5273 & -1.46 & 9.73 & -10.00 & -8.54 & 8.95 & $14.18\ \pm\ 2.14$\\
IC 5332 & -1.95 & 9.68 & -10.07 & -8.12 & 9.30 & $9.01\ \pm\ 0.40$\\
NGC 628 & -1.29 & 10.34 & -10.10 & -8.81 & 9.70 & $9.84\ \pm\ 0.63$\\
NGC 685 & -1.56 & 10.07 & -10.45 & -8.89 & 9.57 & $19.94\ \pm\ 3.01$\\
NGC 1087 & -0.98 & 9.94 & -9.83 & -8.85 & 9.10 & $15.85\ \pm\ 2.22$\\
NGC 1097 & -1.19 & 10.76 & -10.08 & -8.89 & 9.61 & $13.58\ \pm\ 2.05$\\
NGC 1300 & -1.48 & 10.62 & -10.55 & -9.07 & 9.38 & $18.99\ \pm\ 2.86$\\
NGC 1317 & -1.88 & 10.62 & -10.94 & -9.06 & ... & $19.11\ \pm\ 0.85$\\
NGC 1365 & -0.87 & 11.00 & -9.76 & -8.89 & 9.94 & $19.57\ \pm\ 0.78$\\
NGC 1385 & -0.97 & 9.98 & -9.98 & -8.69 & 9.19 & $17.22\ \pm\ 2.60$\\
NGC 1433 & -1.76 & 10.87 & -10.82 & -9.06 & 9.40 & $18.63\ \pm\ 1.84$\\
NGC 1511 & -1.06 & 9.92 & -9.57 & -8.51 & 9.57 & $15.28\ \pm\ 2.26$\\
NGC 1512 & -1.82 & 10.72 & -10.61 & -8.79 & 9.88 & $18.83\ \pm\ 1.86$\\
NGC 1546 & -1.29 & 10.37 & -10.45 & -9.16 & 8.68 & $17.69\ \pm\ 2.02$\\
NGC 1559 & -1.07 & 10.37 & -9.77 & -8.70 & 9.52 & $19.44\ \pm\ 0.45$\\
NGC 1566 & -1.26 & 10.79 & -10.13 & -8.87 & 9.80 & $17.69\ \pm\ 2.02$\\
NGC 1637 & -1.33 & 9.95 & -10.15 & -8.82 & 9.20 & $11.70\ \pm\ 1.01$\\
NGC 1672 & -1.04 & 10.73 & -9.85 & -8.81 & 10.21 & $19.40\ \pm\ 2.93$\\
NGC 1792 & -1.03 & 10.62 & -10.05 & -9.02 & 9.25 & $16.20\ \pm\ 2.44$\\
NGC 1809 & -1.64 & 9.77 & -9.01 & -7.37 & 9.60 & $19.95\ \pm\ 5.63$\\
NGC 2090 & -1.73 & 10.04 & -10.43 & -8.70 & 9.37 & $11.75\ \pm\ 0.84$\\
NGC 2283 & -1.56 & 9.89 & -10.17 & -8.61 & 9.70 & $13.68\ \pm\ 2.06$\\
NGC 2566 & -1.01 & 10.71 & -9.78 & -8.77 & 9.37 & $23.44\ \pm\ 3.53$\\
NGC 2775 & -2.03 & 11.07 & -11.13 & -9.10 & 8.65 & $23.15\ \pm\ 3.49$\\
NGC 2835 & -1.65 & 10.00 & -9.90 & -8.25 & 9.48 & $12.22\ \pm\ 0.93$\\
NGC 2903 & -1.24 & 10.64 & -10.15 & -8.91 & 9.54 & $10.00\ \pm\ 1.99$\\
NGC 2997 & -1.12 & 10.73 & -10.09 & -8.97 & 9.86 & $14.06\ \pm\ 2.80$\\
NGC 3059 & -1.15 & 10.38 & -10.00 & -8.85 & 9.75 & $20.23\ \pm\ 4.04$\\
NGC 3137 & -1.64 & 9.88 & -10.18 & -8.54 & 9.68 & $16.37\ \pm\ 2.34$\\
NGC 3239 & -1.92 & 9.18 & -9.59 & -7.67 & 9.16 & $10.86\ \pm\ 1.05$\\
NGC 3351 & -1.60 & 10.37 & -10.25 & -8.65 & 8.93 & $9.96\ \pm\ 0.33$\\
NGC 3507 & -1.42 & 10.40 & -10.40 & -8.98 & 9.32 & $23.55\ \pm\ 3.99$\\
NGC 3511 & -1.24 & 10.03 & -10.12 & -8.88 & 9.37 & $13.94\ \pm\ 2.10$\\
NGC 3521 & -1.41 & 11.03 & -10.46 & -9.05 & 9.83 & $13.24\ \pm\ 1.96$\\
NGC 3596 & -1.21 & 9.66 & -10.18 & -8.97 & 8.85 & $11.30\ \pm\ 1.03$\\
NGC 3621 & -1.29 & 10.06 & -10.06 & -8.77 & 9.66 & $7.06\ \pm\ 0.28$\\
NGC 3626 & -2.07 & 10.46 & -11.14 & -9.07 & 8.89 & $20.05\ \pm\ 2.34$\\
NGC 3627 & -1.22 & 10.84 & -10.25 & -9.03 & 9.09 & $11.32\ \pm\ 0.48$\\
NGC 4207 & -1.37 & 9.72 & -10.44 & -9.07 & 8.58 & $15.78\ \pm\ 2.34$\\
NGC 4254 & -0.85 & 10.42 & -9.93 & -9.08 & 9.48 & $13.10\ \pm\ 2.01$\\
NGC 4293 & -1.76 & 10.52 & -10.82 & -9.06 & 7.76 & $15.76\ \pm\ 2.38$\\
NGC 4298 & -1.14 & 10.04 & -10.38 & -9.24 & 8.87 & $14.92\ \pm\ 1.36$\\
NGC 4303 & -0.87 & 10.51 & -9.78 & -8.91 & 9.67 & $16.99\ \pm\ 3.02$\\
NGC 4321 & -1.09 & 10.75 & -10.20 & -9.11 & 9.43 & $15.21\ \pm\ 0.50$\\
NGC 4424 & -1.70 & 9.93 & -10.46 & -8.76 & 8.30 & $16.20\ \pm\ 0.69$\\
NGC 4457 & -1.57 & 10.42 & -10.94 & -9.37 & 8.36 & $15.10\ \pm\ 2.00$\\
NGC 4496A & -1.36 & 9.55 & -9.76 & -8.4 & 9.24 & $14.86\ \pm\ 1.06$\\
NGC 4535 & -1.29 & 10.54 & -10.20 & -8.91 & 9.56 & $15.77\ \pm\ 0.37$\\
NGC 4536 & -1.14 & 10.40 & -9.87 & -8.73 & 9.54 & $16.25\ \pm\ 1.12$\\
NGC 4540 & -1.46 & 9.97 & -10.57 & -9.11 & 8.44 & $15.76\ \pm\ 2.38$\\ \hline
    \end{tabular}}
    \label{tab:003}
    \begin{tabnote}
\hangindent6pt\noindent
\hbox to6pt{\footnotemark[$*$]\hss}\unskip%
Columns (2), (5) and (6) were taken from ~\citet{Leroy2021b}. Columns (1), (3), and (4) were calculated using the data. $M(\mathrm{H_2})$ was calculated using $M(\mathrm{H_2}) = \alpha_{\mathrm{co}} L_{\mathrm{co}}$ , Here $\alpha_{\mathrm{co}}=4.35\ M_{\odot}(\mathrm{K\ km\  s^{-1}\ pc^2})^{-1}$ ~\citep{Bolatto2013}.
    \end{tabnote}
\end{table*}

\begin{table*}
    \setcounter{table}{2}
    \centering
    \tbl{(Continued)\footnotemark[$*$]}{%
    \begin{tabular}{p{18mm}ccccccc}
galaxy id & $\log_{10}({M_\mathrm{H_2}/M_*})$ & $\log_{10}({M_{*}}/{M_\odot})$ & $\log_{10}\mathrm{sSFR}(\mathrm{yr^{-1}})$ & $\log_{10}\mathrm{SFE\mathrm({yr^{-1}}})$ & $\log_{10}({M_\mathrm{H_I}}/{M_{\odot}})$ & $d$ (Mpc)\\
 &(1)&(2)&(3)&(4)&(5)&(6) \\
\hline
NGC 4548 & -1.90 & 10.70 & -10.98 & -8.52 & 8.84 & $16.22\ \pm\ 0.38$\\
NGC 4569 & -1.36 & 10.81 & -10.69 & -9.33 & 8.84 & $15.76\ \pm\ 2.38$\\
NGC 4571 & -1.58 & 10.10 & -10.64 & -9.06 & 8.70 & $14.90\ \pm\ 1.07$\\
NGC 4579 & -1.72 & 11.15 & -10.82 & -9.10 & 9.02 & $21.00\ \pm\ 2.03$\\
NGC 4654 & -1.09 & 10.57 & -9.99 & -8.90 & 9.75 & $21.98\ \pm\ 1.14$\\
NGC 4689 & -1.38 & 10.24 & -10.63 & -9.25 & 8.54 & $15.00\ \pm\ 2.26$\\
NGC 4694 & -1.85 & 9.90 & -10.71 & -8.86 & 8.51 & $15.76\ \pm\ 2.38$\\
NGC 4731 & -1.57 & 9.50 & -9.72 & -8.15 & 9.44 & $13.28\ \pm\ 2.11$\\
NGC 4781 & -1.18 & 9.64 & -9.96 & -8.78 & 8.94  & $11.31\ \pm\ 1.18$\\
NGC 4826 & -1.81 & 10.24 & -10.93 & -9.12 & 8.26 & $4.41\ \pm\ 0.19$\\
NGC 4941 & -1.74 & 10.18 & -10.53 & -8.79 & 8.49 & $15.00\ \pm\ 5.00$\\
NGC 4951 & -1.50 & 9.79 & -10.25 & -8.75 & 9.21 & $15.00\ \pm\ 4.19$\\
NGC 5042 & -1.57 & 9.90 & -10.12 & -8.55 & 9.29 & $16.78\ \pm\ 2.53$\\
NGC 5068 & -1.51 & 9.41 & -9.97 & -8.46 & 8.82 & $5.20\ \pm\ 0.22$\\
NGC 5134 & -1.79 & 10.41 & -10.75 & -8.96 & 8.92 & $15.76\ \pm\ 2.38$\\
NGC 5236 & -1.05 & 10.53 & -9.91 & -8.86 & 9.98 & $4.89\ \pm\ 0.18$\\
NGC 5248 & -1.00 & 10.41 & -10.05 & -9.05 & 9.50 & $14.87\ \pm\ 1.32$\\
NGC 5530 & -1.55 & 10.08 & -10.56 & -9.01 & 9.11 & $12.27\ \pm\ 1.85$\\
NGC 5643 & -1.14 & 10.34 & -9.93 & -8.79 & 9.12 & $12.68\ \pm\ 0.54$\\
NGC 6300 & -1.37 & 10.47 & -10.18 & -8.81 & 9.13 & $11.58\ \pm\ 1.75$\\
NGC 7456 & -1.88 & 9.65 & -10.08 & -8.20 & 9.28 & $15.70\ \pm\ 2.33$\\
NGC 7496 & -1.03 & 10.00 & -9.65 & -8.62 & 9.07 & $18.72\ \pm\ 2.82$\\ \hline
      \end{tabular}}
    \label{tab:003}
    \begin{tabnote}
\hangindent6pt\noindent
\hbox to6pt{\footnotemark[$*$]\hss}\unskip%
Columns (2), (5) and (6) were taken from ~\citet{Leroy2021b}. Columns (1), (3), and (4) were calculated using the data. $M(\mathrm{H_2})$ was calculated using $M(\mathrm{H_2}) = \alpha_{\mathrm{co}} L_{\mathrm{co}}$ , Here $\alpha_{\mathrm{co}}=4.35\ M_{\odot}(\mathrm{K\ km\  s^{-1}\ pc^2})^{-1}$ ~\citep{Bolatto2013}.
    \end{tabnote}
\end{table*}

\begin{table*}
    \setcounter{table}{3}
    \centering
    \tbl{Correlation measurements between morphological and physical parameters.\footnotemark[$*$]}{%
    \begin{tabular}{lcc}
Correlation & S$r$ & $p$ \\ \hline
$S$ vs. $A$ & 0.52 & $2\times 10^{-6}$ \\
$C$ vs. $A$ & -0.09 & 0.5 \\
$C$ vs. $S$ & 0.03 & 0.8 \\
$\log_{10}{M_*}$ vs. $C$ & 0.42 & $2\times 10^{-4}$ \\
$\log_{10}{M_*}$ vs. $A$ & -0.21 & 0.07 \\
$\log_{10}{M_*}$ vs. $S$ & -0.32 & $6\times 10^{-3}$ \\
$\log_{10}({M_\mathrm{H_2}/M_{*}})$ vs. $C$ & 0.24 & 0.04 \\
$\log_{10}({M_\mathrm{H_2}/M_{*}})$ vs. $A$ & -0.03 & 0.8 \\
$\log_{10}({M_\mathrm{H_2}/M_{*}})$ vs. $S$ & -0.42 & $2\times 10^{-4}$ \\
$\log_{10}\mathrm{sSFR}$ vs. $C$ & -0.03 & 0.9 \\
$\log_{10}\mathrm{sSFR}$ vs. $A$ & 0.30 & 0.01 \\
$\log_{10}\mathrm{sSFR}$ vs. $S$ & -0.13 & 0.3 \\
$\log_{10}\mathrm{SFE}$ vs. $C$ & -0.07 & 0.5 \\
$\log_{10}\mathrm{SFE}$ vs. $A$ & 0.45 & $7\times 10^{-5}$ \\
$\log_{10}\mathrm{SFE}$ vs. $S$ & 0.33 & $5\times 10^{-3}$ \\
 \hline
    \end{tabular}}
    \label{tab:004}
    \begin{tabnote}
\hangindent6pt\noindent
\hbox to6pt{\footnotemark[$*$]\hss}\unskip%
 Spearman’s rank correlation coefficients (S$r$) and their associated $p$ -values. See Table~\ref{tab:003} for $\log_{10}M_*$, $\log_{10}(M_\mathrm{H_2}/M_{*})$, $\log_{10}\mathrm{sSFR}$ and $\log_{10}\mathrm{SFE}$.
    \end{tabnote}
\end{table*}

\clearpage

\clearpage

\clearpage

\clearpage

\begin{ack}
We thank the anonymous referee for constructive and invaluable comments that improved this paper significantly.
This paper makes use of the following ALMA data,
which have been processed as part of the PHANGS-ALMA CO~(\textit{J}\ =\ 2--1) survey:\par
\noindent
ADS/JAO.ALMA\#2012.1.00650.S,\par
\noindent
ADS/JAO.ALMA\#2013.1.00803.S,\par
\noindent
ADS/JAO.ALMA\#2013.1.01161.S,\par
\noindent
ADS/JAO.ALMA\#2015.1.00121.S,\par
\noindent
ADS/JAO.ALMA\#2015.1.00782.S,\par
\noindent
ADS/JAO.ALMA\#2015.1.00925.S,\par
\noindent
ADS/JAO.ALMA\#2015.1.00956.S,\par
\noindent
ADS/JAO.ALMA\#2016.1.00386.S,\par
\noindent
ADS/JAO.ALMA\#2017.1.00392.S,\par
\noindent
ADS/JAO.ALMA\#2017.1.00766.S,\par
\noindent
ADS/JAO.ALMA\#2017.1.00886.L,\par
\noindent
ADS/JAO.ALMA\#2018.1.00484.S,\par
\noindent
ADS/JAO.ALMA\#2018.1.01321.S,\par
\noindent
ADS/JAO.ALMA\#2018.1.01651.S,\par
\noindent
ADS/JAO.ALMA\#2018.A.00062.S,\par
\noindent
ADS/JAO.ALMA\#2019.1.01235.S,\par
\noindent
ADS/JAO.ALMA\#2019.2.00129.S,\par
\noindent
ALMA is a partnership of ESO (representing its member states), NSF (USA) and NINS (Japan), together with NRC (Canada), MOST and ASIAA (Taiwan), and KASI (Republic of Korea), in cooperation with the Republic of Chile. The Joint ALMA Observatory is operated by ESO, AUI/NRAO and NAOJ. The Joint ALMA Observatory is operated by ESO, AUI/NRAO, and NAOJ. The National Radio Astronomy Observatory is a facility of the National Science Foundation operated under cooperative agreement by Associated Universities, Inc.

\end{ack}

\end{document}